\documentclass[12pt]{article}

\usepackage{scicite}
\usepackage{graphicx}
\usepackage{times}
\usepackage{float}
\usepackage{hyperref}

\newenvironment{sciabstract}{%
\begin{quote} \bf}
{\end{quote}}

\title{Rapid model-guided design of organ-scale synthetic vasculature for biomanufacturing}

\author
{Zachary A. Sexton$^{1}$, Andrew R. Hudson$^{2}$, \\
Jessica E. Herrmann$^{3}$, Dan J. Shiwarski$^{2}$, \\
Jonathan Pham$^{4}$, Jason M. Szafron$^{5,9}$, Sean M. Wu$^{5,6}$, \\
Mark Skylar-Scott$^{1,7,8}$, Adam W. Feinberg$^{2,9}$, Alison Marsden$^{1,10,11\ast}$\\
\\
\normalsize{$^{1}$Department of Bioengineering, Stanford University, Stanford, CA, USA}\\
\normalsize{$^{2}$Department of Biomedical Engineering}\\
\normalsize{Carnegie Mellon University, Pittsburgh, PA, USA}\\
\normalsize{$^{3}$School of Medicine, Stanford University, Stanford, CA, USA}\\
\normalsize{$^{4}$Department of Mechanical Engineering, Stanford University, Stanford, CA, USA}\\
\normalsize{$^{5}$Cardiovascular Institute, Stanford University, Stanford, CA, USA}\\
\normalsize{$^{6}$Division of Cardiovascular Medicine, Department of Medicine}\\
\normalsize{Stanford University, Stanford, CA, USA}\\
\normalsize{$^{7}$Basic Science and Engineering Initiative}\\ 
\normalsize{Children's Heart Center, Stanford University, Stanford, CA, USA}\\
\normalsize{$^{8}$Chan Zuckerberg Biohub, San Francisco, CA, USA}\\
\normalsize{$^{9}$Department of Materials Science and Engineering}\\
\normalsize{Carnegie Mellon University, Pittsburgh, PA, USA}\\
\normalsize{$^{10}$Department of Pediatrics, Stanford University, Stanford, CA, USA}\\
\normalsize{$^{11}$Institute of Computational and Mathematical Engineering}\\
\normalsize{Stanford University, CA, USA}\\
\\
\normalsize{$^\ast$To whom correspondence should be addressed; E-mail:  amarsden@stanford.edu.}
}

\date{\today}

\clearpage

\begin{document} 


\baselineskip14pt


\maketitle 

\section*{One Sentence Summary}
We present an rapid, open-source algorithm capable of generating organ-scale synthetic vascular and multifidelity digital twin models for coupled bio-fabrication and hemodynamic analyses in tissue engineering applications. 

\section*{Abstract}
\begin{sciabstract}
Our ability to produce human-scale bio-manufactured organs is critically limited by the need for vascularization and perfusion. For tissues of variable size and shape, including arbitrarily complex geometries, designing and printing vasculature capable of adequate perfusion has posed a major hurdle. Here, we introduce a model-driven design pipeline combining accelerated optimization methods for fast synthetic vascular tree generation and computational hemodynamics models. We demonstrate rapid generation, simulation, and 3D printing of synthetic vasculature in complex geometries, from small tissue constructs to organ scale networks. We introduce key algorithmic advances that all together accelerate synthetic vascular generation by more than 230-fold compared to standard methods and enable their use in arbitrarily complex shapes through localized implicit functions. Furthermore, we provide techniques for joining vascular trees into watertight networks suitable for hemodynamic CFD and 3D fabrication. We demonstrate that organ-scale vascular network models can be generated in silico within minutes and can be used to perfuse engineered and anatomic models including a bioreactor, annulus, bi-ventricular heart, and gyrus. We further show that this flexible pipeline can be applied to two common modes of bioprinting with free-form reversible embedding of suspended hydrogels and writing into soft matter. Our synthetic vascular tree generation pipeline enables rapid, scalable vascular model generation and fluid analysis for bio-manufactured tissues necessary for future scale up and production. 
\end{sciabstract}

\section*{Introduction}

The fabrication of thick, functional tissues and organs depends upon the successful generation and integration of perfusable vascular networks \cite{bib1}. As we look to manufacture living matter beyond microscale dimensions, ensuring adequate oxygen supply, nutrient delivery, and waste removal becomes paramount \cite{bib2,bib3}. Natively, cells of most soft, vascularized tissues lie within $100\mu$m of a blood vessel; in metabolically demanding tissues, this distance can be less than $20\mu$m and vessel density may be greater than $2500$ vessels per mm\textsuperscript{3} \cite{bib4,bib5,bib6}. Hierarchical structure allows vascular networks to reach the extensive scales necessary for organ perfusion without sacrificing transport efficiency \cite{bib7,bib8}. Thus, scalability of manufactured tissues intimately depends on recapitulating the tree-like vascular organization observed \textit{in vivo}.

Recent advances in tissue manufacturing allow fabrication of multi-material constructs replete with multiple cell types and extracellular matrix compositions, at resolutions and speeds necessary for viable 3D tissues \cite{bib9,bib10,bib11,bib12}. However a robust method for generating a multi-scale and biomimetic perfusable network within these tissues remains elusive. Though it is possible to segment vasculature directly from image data, these data often do not exist at adequate resolution for arbitrary shapes used in bioprinting. To date, only cryomicrotomes and light sheet microscopy techniques have allowed direct segmentation of the microvasculature \cite{bib13,bib14}. Furthermore, engineered tissues rarely have direct anatomical analogues from which to extract imaged vascular networks. Modern-day perfusion designs have thus relied on simple lattices, which poorly reproduce the topology and hemodynamics of native vasculature \cite{bib15,bib16,bib17,bib18}.  While acceptable for tissue manufacturing at low cell densities, these idealized designs fail to satisfy the metabolic demand of clinically relevant cell densities, thereby precluding in vitro viability of larger-scale manufactured tissues. It is thus clear that a unified and organ-scalable approach for vascular network generation, simulation, and 3D manufacture is needed.

In the past twenty years, cardiovascular computational fluid dynamics (CFD) has emerged as an important tool for patient specific treatment planning, medical device design, and understanding mechanobiological processes of disease progression. Towards these needs, the CFD community developed methods to generate microvascular networks for incorporation into hemodynamic models. Structured trees of fractal topology and morphometric trees have been employed to investigate microvascular autoregulation, to quantify wall shear stresses, and to provide boundary conditions for 3D simulations \cite{bib19,bib20,bib21,bib22}. While these can provide basic hemodynamic information, typically using a Poiseuille flow assumption, they cannot provide the 3D spatial information needed for biofabrication of vascular networks. 

Algorithms to construct vascular networks in 3D volumes  have therefore been proposed and include global constrained optimization (GCO), Monte Carlo, Lindenmayer systems, fractal trees, growth-based optimization (GBO), and constrained constructive optimization (CCO), among which GBO, CCO and GCO exploit hemodynamic principles during vascular generation to ensure sufficient perfusion \cite{bib23,bib24,bib25,bib26,bib27,bib28,GBO}. However, prior CCO methods had several significant drawbacks including long build times (hours to days for complex models), limited handling of non-convex volumes, and an inability to predict resulting hemodynamics.  Although some recent work has extended both CCO and GCO for non-convex perfusion territories, the methods are computationally expensive and CCO remains the only algorithm in this family able to produce closed fluid circuits, a requirement for engineered tissues \cite{bib29,bib30,bib31}. Thus, despite the above limitations, CCO has been widely adopted for iterative vascular generation, with applications spanning both patient-specific CFD and perfusion of 3D printed tissues \cite{bib31,bib32}.

To overcome the above limitations, we introduce a synthetic vascular toolkit that generates organ-scale networks within minutes and provides a complete automated pipeline from model generation to hemodynamic simulation and fabrication. Here we combine CCO methods with recent techniques for CFD hemodynamic simulation and 3D printing to provide matched flow analysis and vascular geometry, generated in minutes, in support of tissue fabrication needs. We extend these methods for complex perfusion territories, multi-fidelity hemodynamic simulation, and closed-loop perfusion networks. We demonstrate our workflow by perfusion of 3D bioprinted vascular networks in a bioreactor and corroborate flow behavior using CFD models. We further characterize and simulate networks within larger engineered designs and organ-scale anatomies to motivate large-scale vascular planning of engineered tissues. 

\section*{Results}
\subsection*{Overview of synthetic vascular data structures}
We introduce a multi-modal data structure fundamental for accelerating synthetic vascular generation (Figure 1). Leveraging flexible and efficient N-dimensional array containers, data is stored hierarchically, mirroring vascular architecture. Single rows correspond to linear vessel elements containing all necessary physical (e.g. position, length, radius) and derived properties (e.g. connectivity, bifurcation ratios, hydraulic resistance) for vascular generation (Figure 1A). 2D arrays form independent, open-loop vascular trees (single-inlet, multiple outlets); 3D arrays define connected trees to create closed-loop, interpenetrating networks. 4D arrays allow for multiple non-interpenetrating networks to generate within the same domain, for example the arterial/venous, and lymphatic networks (Figure 1B). Array structures leverage intuitive vectorization, broadcasting, reduction, and continuous memory operations to accelerate synthetic vascular generation when compared to alternative linked list implementations \cite{numpy} (Figure 1C). 

\begin{figure}[H]
    \centering
    \includegraphics[scale=0.65]{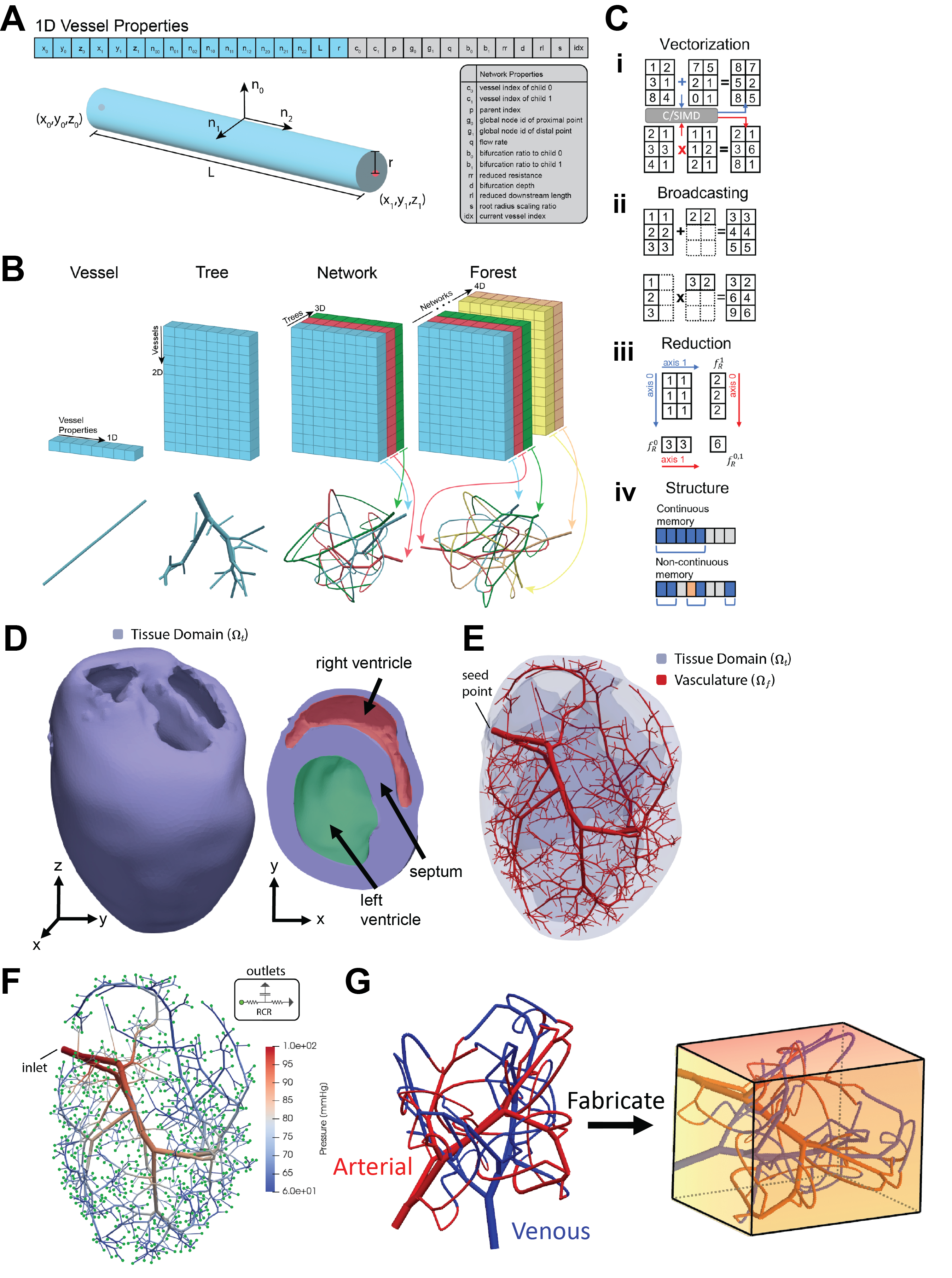}
    \label{fig1}
\end{figure}

\noindent \textbf{Fig. 1. Fundamental data structures and capabilities of synthetic vascular toolkit.} \textbf{(A)} data for linear vessel segments defined as row-wise double-precision elements representing physical (blue) and hierarchical (grey) properties. \textbf{(B)} an overview of dimensional contributions to vascular topology. Although depicted as equal sizes, networks within forests can be of different sizes (e.g. $[:, N, :, 0]$ and $[:,M,:,1]$); however, trees within a given network must be of equal size to ensure closure (e.g. $[:, N, 0, 0]$ and $[:, N, 1, 0]$). \textbf{ (C)} array-specific operations accelerate vascular generation through vectorization, broadcasting, and reduction which leveraging single-instruction multiple data (SIMD) accelerators in modern CPU architectures to perform simultaneous execution of parallel instructions. \textbf{(D)} a biventricle tissue tissue domain ($\Omega_{t}$) defines the space in which synthetic vasculature is generated with challenging thin-walled regions including the annotated ventricular septum and right ventricular wall \textbf{(E)} open-loop synthetic vasculature tree with 1000 outlets defining the fluid domain ($\Omega_{f}$) for blood or media flow. The root location of the tree is denoted by the seed point. \textbf{(F)} a matched CFD simulation automatically generated from the same 1000-outlet vascular tree. The inlet boundary admits a flow waveform while outlets are coupled to downstream circuit elements representing the impedance of capillaries/venous vessels. Normalized pressure is displayed. \textbf{(G)} connections between separate synthetic arterial and venous vascular trees form a close-loop network ideal for biofabrication.
\clearpage

To test the performance of our data structure, we evaluate three common vessel generation subroutines: vessel bifurcation, rescaling, and collision detection (See Supplement Text). During vessel bifurcation, a new vessel is appended to the current tree to reach some new area within the tissue. Determining the optimal branch for appending a new vessel requires iteration through the existing tree. From our choice of array-based data structures earlier, we observe improvement over linked-list iteration used in previous approaches, achieving an 6-fold speed up (Supplement Figure 1A). Next, we evaluate the efficiency of propagating a scaling factor down an existing tree from the root. As synthetic vasculature grows, it accommodates greater flow thus demanding larger vessel diameters. Within sufficiently large vasculature ($\geq500$) terminal segments, a 28-fold improvement from array broadcasting is achieved (Supplement Figure 1B). Last, collision detection achieves the greatest improvement with up to 5-fold acceleration due to a combination of vectorization, broadcasting, and reduction operations (Supplement Figure 1C). Because these subroutines may be called hundreds or thousands of times during vascular generation, modest accelerations can result in significantly improved performance overall. A judicious choice of scalable array data structures allows for efficient implementation of iterative methods necessary for vascular generation with an interpreted language. 

In application, our synthetic vascular toolkit handles arbitrarily complex tissue domains, $\Omega_{t}$, including cavities and thin walls. Figure 1D presents a biventricle tissue domain as an example of a highly non-convex geometry of interest. Here the septum and right ventricle wall are particularly challenging regions for previous synthetic vascular algorithms to traverse efficiently. To avoid such challenges, we use implicit reconstructions of tissue domains to efficiently explore tissue domains during vascular generation. A generated vascular tree within the original biventricle tissue domain is shown which reaches regions of the left and right ventricle walls as well as the ventricular septum (Figure 1E). Here, the vascular domain, $\Omega_{f}$, represents the blood volume of the vasculature. Because hemodynamic boundary conditions are often more complex than the simplified Poiseuille assumptions made during vascular generation, multi-fidelity hemodynamic simulations are automatically coupled to generated vasculature for optional \textit{post hoc} CFD analysis (Figure 1F). The normalized pressure through a synthetic vascular tree is shown in Figure 1F. As expected, pressure uniformly decreases down the vascular tree along bifurcation depth. Comparisons of pressure dissipation across alternative lattice architectures demonstrate more uniform dissipation of flow rates across vessels in synthetic vasculature (Supplement Figure 2). Last, synthetic vasculature can be exported as watertight networks for 3D bio-fabrication using common mesh file formats (such as .stl or .3mf) or direct tool path files (such as .gcode). Figure 1G illustrates a close-loop arterial-venous system for a tissue volume. 

\subsection*{Performance of acceleration techniques during synthetic vascular generation}
Beyond the choice of data structure, we introduce four components to improve efficiency and extend synthetic vascular generation: 1) partial binding optimization, 2) partial implicit volumes, 3) collision avoidance algorithms, and 4) generation of close-loop vasculature. During vascular generation, new vessels are appended by creating bifurcations within existing trees. Bifurcation points are subsequently optimized with respect to a given cost function $\mathcal{C} = w(x)\sum l^{u}_{i}r^{\lambda}_{i}$, $\forall i \in \mathcal{T}$ which is a linear combination of vessel radii, $r_{i}$, and lengths, $l_{i}$, for each vessel in the tree, $\mathcal{T}$, scaled by a penalty function, $w$, dependent on the bifurcation position vector, $x$. During optimization, hydraulic changes back-propagate to the root of synthetic trees which re-scales the entire tree in a forward-propagating pass during each evaluation. Local optimization of bifurcation locations is thus dominated by tree traversal operations. We introduce a partial binding scheme (Figure 2A) whereby intermediate vascular structures are stored locally to avoid repeated tree traversals.
\begin{figure}[H]
    \centering
    \includegraphics[scale=0.52]{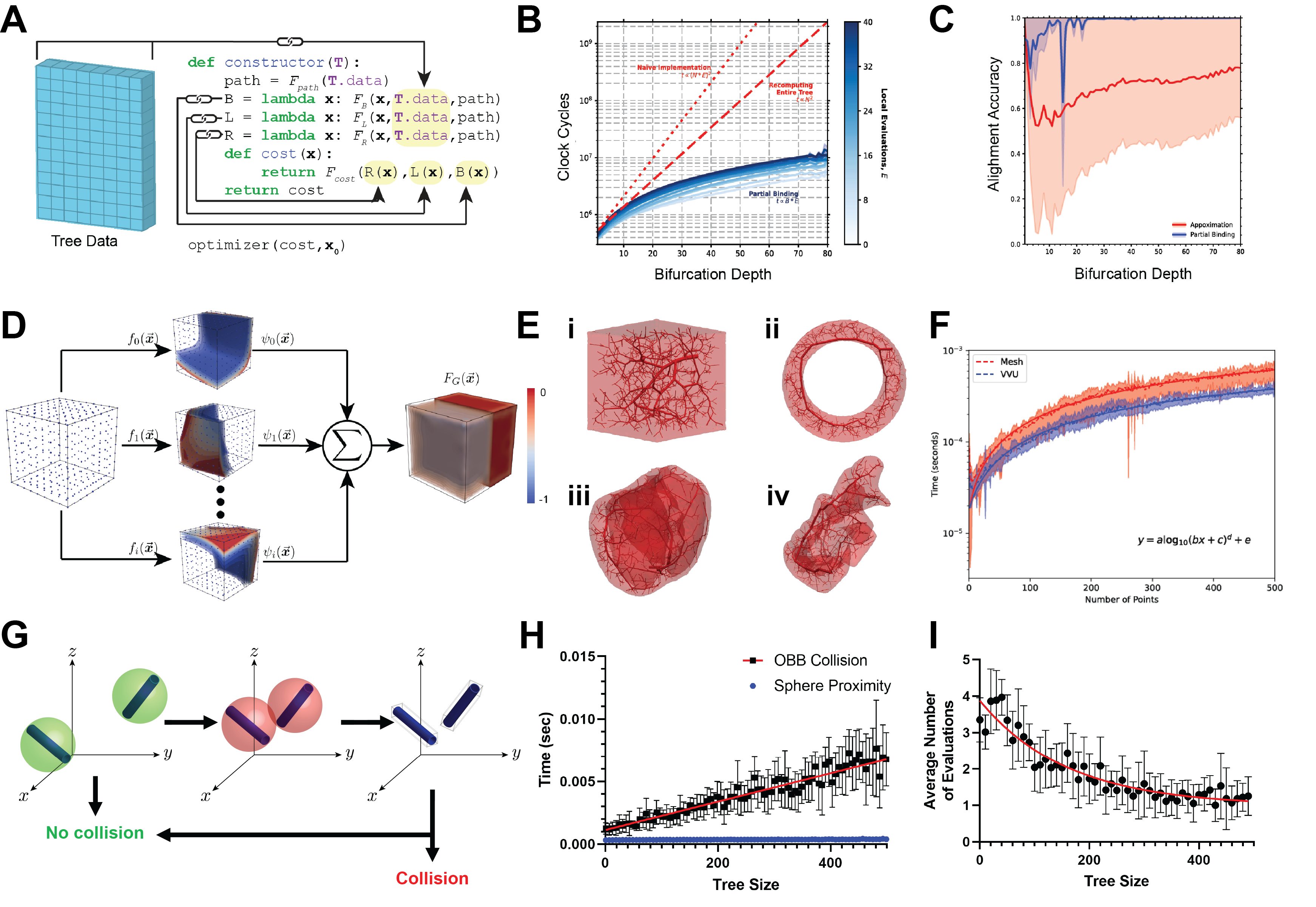}
    \label{fig2}
\end{figure}
\noindent {\bf Fig. 2. Performance of synthetic vascular acceleration techniques.} {\bf (A)} conceptual schematic of partial binding scheme for bifurcation optimization. The outer scope "constructor" builds functions computing the bifurcation ratios ($F_{\beta}$), hydraulic length updates ($F_{L^{*}}$), and hydraulic resistance updates ($F_{R^{*}}$) from the locally bound tree configurations which depend on the bifurcation position vector $x$. The resulting cost function ($F_{cost}$) is returned for optimization. {\bf (B)} theoretic scaling complexities (red) for bifurcation optimization with previously reported schemes compared against achieved scaling complexity of partial binding techniques (blue) reference to bifurcation depth in full, balanced trees. Shaded regions $\pm 2$ SD, $N=100$ {\bf (C)} alignment accuracy against highly-resolved finite difference for partial binding (blue) and previously reported approximation (red) methods. Solid lines represent median alignment performance. Shaded regions span the best and worst performance for bifurcation alignment. $N=100.$ All trees are generated to 8000 outlets within a cube perfusion volume. {\bf (D)} schematic of implicit partial volumes displays the decomposition of surface point data into local patches which are blended together during implicit reconstruction {\bf (E)} implicit partial volumes are reliable for synthetic vascular generation across engineering and anatomic geometries including (i) cube, (ii) annulus, (iii) bi-ventricle heart, and (iv) brain gyrus. Vascular trees with 1000 outlets are shown for each domain {\bf (F)}
scaling complexity of mesh-based and implicit method evaluation times, shaded regions $\pm 2$ SD. $N=25$ {\bf (G)} schematic overview of collision avoidance during tree generation. Sphere proximity detects vessels with potential collisions to be evaluated by more precise oriented bounding boxes (OBB) {\bf (H)} sphere proximity is an inaccurate but cheap alternative to full OBB evaluations and {\bf (I)} when used as a filter, eliminate the majority of expensive OBB collision evaluations as trees increase in complexity. Error bars $\pm 2$ SD. $N=100$ for synthetic trees grown to 500 outlets. 

\clearpage
In previous literature, bifurcation optimization typically required two complete tree traversals during each cost function evaluation resulting in computational scaling of at least $\mathcal{O}(N^2)$, where $N$ is the number of vessels in the tree. As vasculature grows beyond a few hundred vessels, this optimization scheme becomes significantly expensive leading to days of compute time (Figure 2B). Motivated by this inefficiency, recent approaches have employed approximations to the cost function to produce constant time optimization methods \cite{bib31}. While excellent for bifurcations of greater depth, approximation methods tend to have greater disagreement with the true optimum at depths closer to the root vessel (depth $\leq20$). In contrast, partial binding scales at worst $\mathcal{O}(\log(N))$ and evaluates the true cost function; therefore accuracy of the bifurcation position depends only on optimization tolerances. In side-by-side comparison, partial binding achieves median alignment accuracy $\geq60\%$ across all bifurcation depths and $\geq95$ for deep bifurcations (Figure 2C). A selection of optimizers are considered from which brute force searches on semi-coarse grids offer reasonable trade-off between computational expense and accuracy (Supplement Figure 3). 

As realistic tissue geometries are typically non-convex and expensive to parse, we implement a partial implicit volume method to reconstruct arbitrary tissue domains for fast interrogation during vascular generation. This method extends variational point set surface construction as a partition of unity (see Supplement text Methods; Supplement Figure 4A,B). This approach allows for a direct definition of an implicit function from sampled data as a quadratic optimization problem without poor time and space complexities previously reported at $\mathcal{O}(D^{3}N^{3})$ \cite{vpss}. In addition, our approach avoids ill-conditioning of the implicit function (Supplement Figure 4C,D). We tested reconstruction with 10 contrived engineering geometries and over 200 anatomic regions of brain tissue to assess performance across arbitrary domains \cite{anatomic_models}. Time and accuracy results of the implicit partial volume method demonstrate efficient and reasonably accurate reconstructions for synthetic vascular generation (Supplement Table 4; Supplement Figure 5). Comparing specific generation times for 8000 terminal trees within a gyrus geometry against previous demonstrations in literature, we observe a median time of 15 minutes versus 59.96 hours-a more than 230 fold acceleration\cite{dcco}. Shown are two engineered and two anatomic domains including a cube, annulus, heart, and brain gyrus which each successfully generate 1000-outlet synthetic vascular trees within the tissue domains (Figure 2E; Supplement Video 1). Even for sharp surface edges, we notice reasonable recovery of boundary features which can be further improved by adapting the underlying implicit functions (See supplement text; Supplement Figure 6). Importantly, implicit partial volumes provide cheap point enclosure evaluations compared to mesh point-in-polygon methods; even meshes wrapped in hierarchical data structures (e.g. kd-tree, octree) confer more expensive evaluations as the number of mesh elements is typically much greater than the number of implicit functions needed to describe the same domain (Figure 2F; Supplement Figure 7).

Another important accelerator towards efficient vascular generation is the triaging of collision avoidance tests. Since collisions are checked after each newly appended vessel, this subroutine quickly scales in computational cost. We employ unorientable bounding surfaces around each vessel segment as a cheap test for plausible collisions before proceeding with more expensive oriented bounding box (OBB) tests (Figure 2G). As size of synthetic vasculature trees increases, we observe that the majority of potential collisions can be resolved in 1 or 2 OBB tests avoiding another source of poor computational scaling for large-network generation (Figure 2H \&2I). All together the previous acceleration techniques yield efficient synthetic vascular generation for open-loop networks for more than 200 total engineering and anatomic tissue domains (Supplement Figure 8). A selection of synthetic vascular networks and their performance times within twenty brain tissue domains are shown (Supplement Table 5; Supplement Figure 9).

Last, synthetic vasculature trees can be readily connected within arbitrary shapes to form close-loop networks through multi-objective optimization procedures (See supplement text). Connections between vascular terminal segments balance competing objectives including vessel tortuoisity, vessel collisions, boundary violations, and total blood volume that result from manipulating cubic Bezier curves within the tissue domain. Typically entangled vascular models have been limited to generation within convex domains; we demonstrate our approach easily applies across non-convex geometries such as an annulus and biventricle heart to provide a means to actualize perfusion designs with anatomic and bioengineering use cases. Furthermore, connections can readily be made among multiple interpenetrating trees to form complex vascular mixing networks with minimal scripting changes. 

\subsection*{Automatic multifidelity hemodynamics modeling}
Our method to generate watertight synthetic vascular tree models enables us to perform 3D hemodynamics simulations capable of evaluating spatiotemporal dynamics of flow necessary for resolving quantities of interest including local wall shear stress and pulse-wave velocity. Such models better inform design constrains toward desired pressure gradients and shear stress ranges matching physiologic physics of native vasculature. This is an advance over previous methods which commonly represented synthetic vasculature as a collection of discrete cylinders that could not readily be discretized into a computational mesh for simulation. Creating simulation-ready models allows us to predict performance of vascular tree designs in advance of fabrication by assessing perfusion, wall shear stress, and other parameters critical to tissue viability.  

To perform simulations, we developed automated methods to convert discrete vascular tree models into 3D watertight, simulation-ready meshes (Figure 3A). We leveraged the open source pipelines for patient specific modeling and finite element simulation of blood flow available in the open source SimVascular project. In SimVascular, simulations are performed in the following steps: 1) anatomic models are segmented from medical image data by generating centerline paths and segmenting the lumen of all vessels of interest, 2) unstructured meshes are generated to discretize the 3D volume, 3) boundary conditions are applied, 4) hemodynamics simulations are run using the finite element method \cite{updegrove,marsden_multiscale_review}. Unlike traditional CFD model-building frameworks for which many steps of this pipeline are performed manually, we developed methods to automate the modeling pipeline for synthetic vasculature. Thus, for each synthetic vascular tree produced, we perform automated contour checking, smoothing, meshing, and boundary condition assignments (See supplement text), and blood flow simulation. The resulting 3D finite element simulation is parallelized and run on high-performance computing clusters to obtain resolved, high-fidelity hemodynamic quantities of interest (see supplement text; Figure 3B). Normalized pressure, velocity streamlines, and wall shear stress are presented for a steady-flow simulation to highlight the integrity of this automatic pipeline; time-varying flow is also simulated using idealized pulsatile waveforms \cite{kassab}. 

\begin{figure}[H]
    \centering
    \includegraphics[scale=0.55]{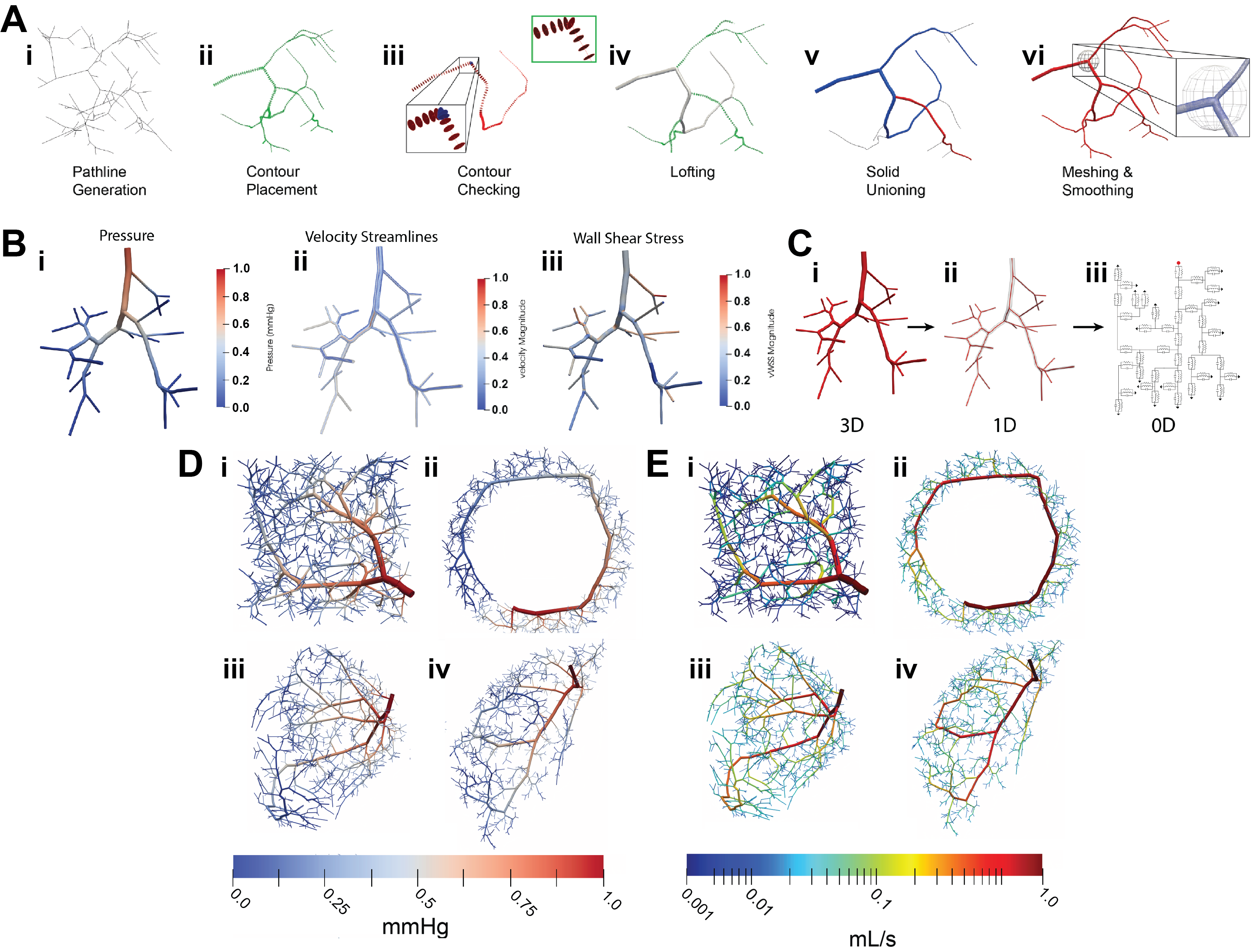}
    \label{fig3}
\end{figure}
\noindent {\bf Fig. 3. Automatic multifidelity hemodynamics modeling.} {\bf (A)} major steps for automatic 3D watertight model generation from discrete synthetic vessels {\bf (B)} normalized pressure, velocity streamlines, and wall shear stress obtained from 3D FEM simulation with a steady inflow profile {\bf (C)} illustration of reduced order 1D and 0D models extracted from original 3D model files ({\bf D} and {\bf E}) average pressure and flow rates obtained from 0D steady-flow simulations for vascular trees with 1000 terminals within (i) cube, (ii) annulus, (iii) bi-ventricle, and (iv) brain gyrus tissue domains. 
\clearpage 

Depending on the caliber and number of vessels, high-fidelity 3D FE simulations may be too expensive, or it may be unnecessary to fully resolve hemodynamic quantities at all levels of the tree. Synthetic vasculature easily extends to hundreds or thousands of vessels where a majority exist in laminar or near-Stokes flow regimes (Re $\leq 50$). These low Reynolds number flows are well described by simplified, or low-fidelity, simulations \cite{reduced_models}. Namely, $1$D and $0$D simulations allow for investigation of large vessel networks in a computationally efficient manner. $1$D simulations capture only one spatial dimension by integrating the NS equations over the vessel cross-section (see supplement methods 1D). $0$D simulations further simplify the governing fluid equations by making a circuit analogy to produce a set of inexpensive, differential algebraic equations (see supplement methods 0D) which can be solved on most personal computers. Such lower fidelity models have been shown to produce accurate flow and pressure waveforms, particularly in healthy vasculature at low Reynold's number. Here, we demonstrate an automated pipeline to extract low fidelity simulation inputs from the original 3D models (Figure 3B) and run either 0D or 1D models. We demonstrate these low-fidelity simulations on extensive vascular networks for the same cube, annulus, heart, and brain gyrus as presented in Figure 2E. We compute bulk flow quantities, including average pressure and volumetric flow rate, with a 0D simulation over a cardiac cycle (Figure 3C,D). Unlike the simple assumptions underpinning vascular generation, vessel resistance and compliance are well captured for time-varying flow magnitude and pressure pulse propagation down vascular trees (Supplement Video 2). Additionally, wall shear stress is well-approximated at these scales from idealized assumptions of pipe flow, since these forces are well-known to affect vascular remodeling (see methods 0D analysis).

\subsection*{3D biofabrication with on-demand synthetic vasculature designs}
We employed two methods to demonstrate 3D printing of vascular networks generated with our model-driven pipeline. We first applied freeform reversible embedding of suspended hydrogels (FRESH; see FRESH printing methods) to produce a simple planar network with four vessels to ensure that vascular models are indeed watertight and adequate for perfusion. FRESH is an embedded 3D printing method in which gelling materials are printed into a non-Newtonian Bingham plastic support material derived from gelatin microparticles compacted within a fluid phase. The shear-dependant transition between solid-like and liquid-like behaviors prevents deformation of the extruded ink and allows needle movement through the bath. Above 37$^{\circ}$C gelatin supports can be melted to allow for easy evacuation of the vascular channels (Figure 4A). This process yields a high-fidelity collagen print closely matching the generated model geometry as assessed through optical coherence tomography obtained during printing (Figure 4B). We observe minimal deviation ($\leq 100 \mu$m) between the printed model and the original after computing registration and error gauging throughout vascular channels of the print. After removal of support material, the presence of the watertight lumens was verified by an advancing dye front within the planar network (Figure 4C; Supplement Video 3). Print resolution is constrained by a 150$\mu$m extrusion needle diameter and $60\mu$m layer height which is below our smallest feature diameter of 400$\mu$m. In a matched model, we perform a 3D CFD simulation using the same inlet $0.25$ mL/min flow rates for the advancing dye front to characterize lumen stresses from pressure and wall shear stress (Figure 4D). 

\begin{figure}[H]
    \centering
    \includegraphics[scale=0.45]{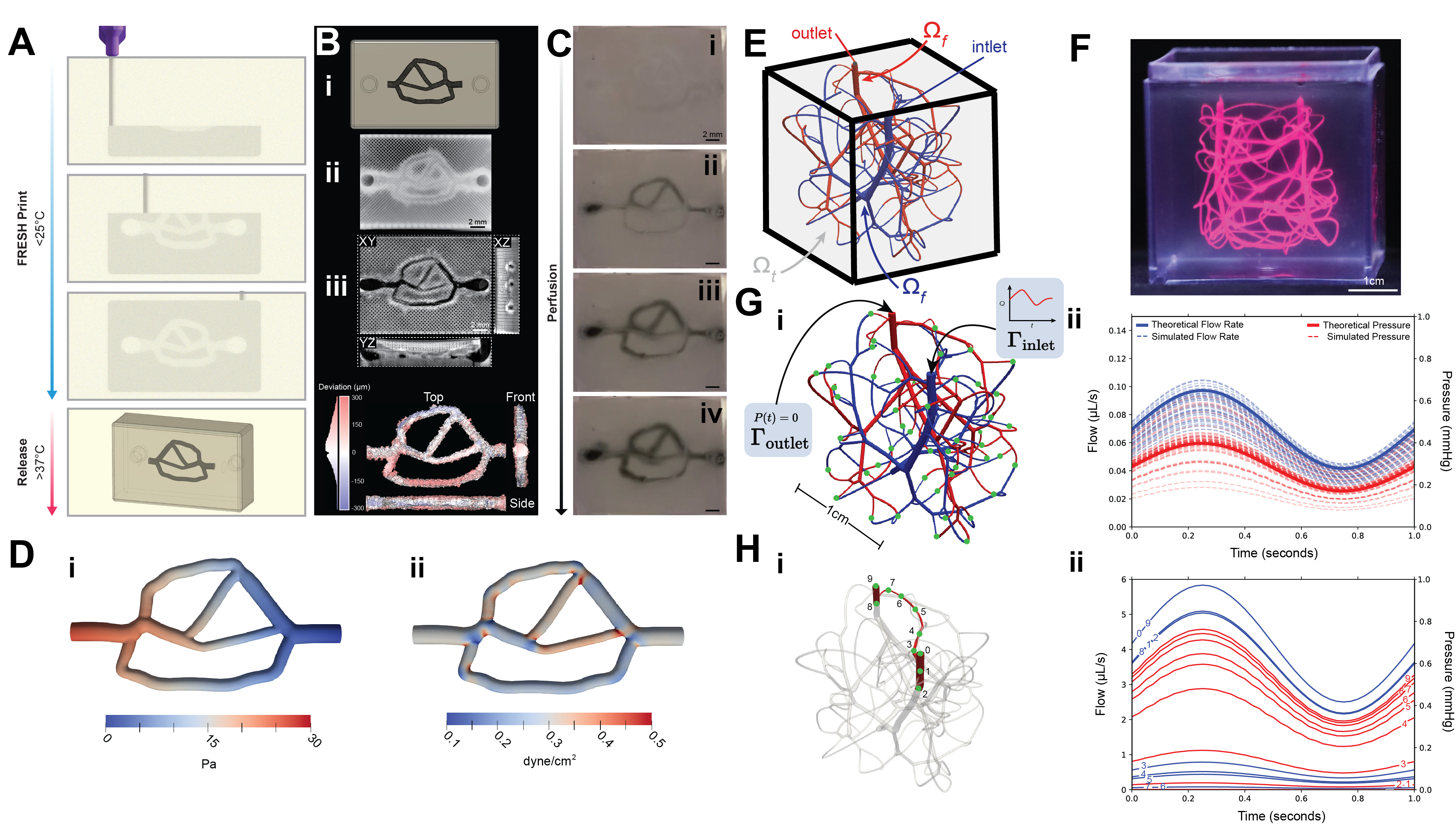}
    \label{fig4}
\end{figure}
\noindent {\bf Fig. 4. Synthetic vascular applications in 3D biofabrication.} {\bf (A)} FRESH printing process proceeds in two steps 1) layer-by-layer printing of bioinks into a cool thermoreversible sacrificial support ($\leq 25^\circ$C) followed by 2) melting and removal of sacrificial material upon heating ($\geq 37^\circ$C) {\bf (B)} OCT imaging of FRESH printing construct {\bf (C)} perfusion of vasculature with black dye verifying physical integrity of printed lumens {\bf (D)} 3D CFD simulation of steady perfusion into the planar network (Q=0.25mL/min) with (i) pressure and (ii) wall shear stress distributions obtained throughout the network {\bf (E)} idealized model of 3d network with annotated inlet, outlet, and cube tissue domain $\Omega_{t}$ and fluid domain $\Omega_{f}${\bf (F)} image of printed 3d network using extruded soft matter {\bf (G)} 3D CFD pulsatile flow simulation for printed network with mean flow rate (Q=0.25mL/min) and amplitude of 0.1mL/min. (i) Annotated points depict midpoints of vessels within the network. (ii) simulated flow rate and pressure waveforms at vessel midpoints (dashed lines) are shown along with idealized uniform theoretical flow and pressure waves (solid lines) {\bf (H)} (i) 10 annotated points represented sampled cross-sections along a single vessel (ii) flow rate and pressure waveforms along sampled points show decreasing flow rates 0 $\rightarrow$ 7 followed by compensate increases from 7 $\rightarrow$ 9. Pressure waveforms attenuate from inlet to outlet 0 $\rightarrow$ 9. 
\clearpage

To demonstrate the utility of this model-driven pipeline for tissue fabrication, a non-planar single-inlet, single-outlet vascular network was created and printed using extrusion embedded printing. In this example the vascular domain ($\Omega_{f}$) was created within a cube tissue domain ($\Omega_{t}$) (Figure 4E). Fabrication was accomplished on a scaled model through embedded extrusion printing of a granular ink into a granular support bath (carbopol-in-carbopol) to demonstrate qualitative viability of vascular model features with soft matter (see methods). An important feature of generated synthetic vasculature is the equivalent hydraulic resistance across all branches. Such a characteristic implies uniform perfusion throughout tissue and commensurate improvements in long-term tissue viability. Prior to printing, we simulated steady and pulsatile flow using digital-twins of the proposed networks to assess hemodynamic integrity (Figure 4G,H; Supplement Figure 8). Flow media for all simulations we assume the density and viscosity of the working fluid approximates whole blood ($\rho = 1.06$ g/cm$^3$, $\nu = 0.04$cP). At the inlet, the volumetric flow rate was idealized by a sinusoid with mean set to equal the previous steady flow rate with an amplitude of 0.1mL/min. We applied the same zero-pressure outlet boundary condition. To analyze the simulation results, we first verified that flow and pressure waveforms maintained similar profiles relative to ideal theoretical flow and pressure distributions at all of the mid-points of the connecting vessels (Figure 4Gi-ii). We obtained the simulated time-varying flow and pressure waves at each of the selected midpoints. We find that both flow and pressure at the midpoints remain roughly similar to ideal theoretical distributions with absolute deviations of 0.02$\mu$L/min and 0.03mmHg for flow and pressure, respectively. These differences are most likely due to smoothing of vessels from discrete cylinder networks; future work will be needed to adjust vessel diameters from higher-fidelity CFD simulations to compensate vascular topology for greater flow uniformity. We similarly show within single vessels that pressure waves attenuate along vessel lengths as fluid progresses from the inlet (Point 0) to the outlet (Point 9) while flow rate decreases toward the halfway point (Points 6 \& 7) before returning to the inlet flow value at the outlet (Figure 4Hi-ii). Such simulations agree with intuitions of viscous power losses and mass conservation assumed for the perfusion networks which are prerequisite for future simulations of perfusion and stress propagation within the tissue domain. 

\section*{Discussion and Conclusion}
The future of engineered tissues depends on a robust software infrastructure capable of designing and validating vascular networks through physics-based, functional targets. While simple perfusion architectures adequately serve as playground tests for 3D fabrication methods, their over-reliance limits real-world applications and risks over-fitting fabrication methods to simple design solutions. The need to generate custom vasculature across multiple scales has posed a significant barrier to progress in the bioprinting community. Here we present a comprehensive model-driven pipeline for efficient vascular generation in arbitrary geometries capable of creating watertight models suitable for biofabrication and digital-twins for advanced multifidelity hemodynamic simulations. Long-term, engineered tissues aspire to replace damaged or diseases tissues in patient-specific interventions. Validating such tissues will require simulation-optimized designs prior to resource intensive fabrication. Like surgical planning, numerical models will become increasingly necessary to generate rational designs and predict functional efficacy of engineered tissue prior to clinical translation \cite{marsden_review}. With the presented model-driven pipeline, simple script commands yield complex models intended for wide-ranging applications in biofabrication and patient-specific modeling. Using accelerated techniques in iterative vascular generation and implicit reconstructions of tissue domains, we create complex vascular models within minutes--opening up the potential for more extensive hemodyanmic analysis and design optimization in future synthetic vascular structures. We further demonstrate the fabrication of synthetic vascular models using well-known bioprinting methodologies to illustrate utility in real-world in vitro applications. By combining our ability to automate a model-guided design of organ-scale vascular networks, and recent work on the billion cell-scale production of organ specific cell types, we believe that the practice of organ engineering is coming within reach \cite{billion_cells}. 

\bibliography{scibib}

\bibliographystyle{Science}

\section*{Acknowledgments}

The project described was supported, in part, by the Additional Ventures Cures Collaborative and the Stanford Cardiovascular Institute. Z.A.S. was funded by the National Science Foundation Graduate Research Fellowship Program (DGE-1656518) and the Stanford Graduate EDGE Fellowship. J.E.H was funded by the Berg Scholars Program and, in part, by the Alpha Omega Alpha Carolyn L. Kuckein Student Research Fellowship, Dorothy Dee and Marjorie Helene Boring Trust Award, and Medical Scholars Research Program. D.J.S. was funded by the National Heart, Lung, and Blood Institute of the National Institutes of Health under Award Number K99HL155777. J.P. was funded by NIH grant R01EB029362. J.M.S. acknowledges funding support from the Parker B. Francis Fellowship. S.M.W was supported from (American Heart Association - Established Investigator Award, Hoffmann/Schroepfer Foundation, Additional Venture Foundation, Joan and Sanford I. Weill Scholar Fund, and the NSF RECODE grant. M.S. was funded by the National Heart, Lung, And Blood Institute of the National Institutes of Health under Award Number DP2HL168563. A.L.M. reports funding from grant 5R01HL14171205, NIH SCH grant 5R01EB02936204, and NSF SimCardio grant NSF1663671. 

\section*{Code Availability}
All synthetic vascular library source code is available in the GitHub repository,\newline \url{https://github.com/zasexton/Synthetic-Vascular-Toolkit}. The project package is also available as a ready-to-use python module which can be pip installed from the Python Package Index, \url{https://pypi.org/project/svcco/}.

\section*{Supplementary materials}
Materials and Methods\\
Supplementary Text\\
Figs. S1 to S9\\
Tables S1 to S5\\
Videos S1 to S3\\
References \textit{(43-77)}
\end{document}


\noindent \textbf{Supplement Materials: Rapid model-guided design of organ-scale synthetic vasculature for biomanufacturing}

\baselineskip14pt



\noindent \textbf{The PDF File includes}\\
Materials and Methods \newline 
Supplementary Text \newline
Figs. S1 to S9 \newline
Tables S1 to S5 \newline
Supplementary References

\noindent \textbf{Other Supplementary Materials for this manuscript include the following}\newline
Movies S1 to S3

\clearpage
\begin{table}[h]
\begin{center}
\begin{minipage}{300pt}
\caption{Vascular Generation Symbols and Constants}\label{Vascular Generation Symbols Table}%
\begin{tabular}{@{}llll@{}}
\toprule
Symbol & Description & Units & Value\\
\midrule
$\mathcal{T}$& vascular tree        &                            &  \\
$Q$          & flow rate            & mL/s                  &  \\
$\Delta P$   & pressure difference  & mL/s                  &  \\
$l$          & length               & cm                              &  \\
$r$          & radius               & cm                              &  \\
$\mu$        & dynamic viscosity    & cP                              & 4\\
$R$          & hydraulic resistance & $\text{dyne}/\text{cm}^5$             &    \\
$R^{\ast}$   & reduced resistance   & $\text{dyne}\cdot \text{cm}^{3}$              &    \\
$\gamma$     & scaling exponent     &                             & 3  \\
$\beta$      & bifurcation ratio    &                             &    \\
$\alpha_r$   & radius exponent      &                             & 2  \\
$\alpha_l$   & length exponent      &                             & 1  \\
$\zeta$      & weighting function   &                             &    \\
$C_\mathcal{T}$& tree cost          & $\text{cm}^{\alpha_r+\alpha_l}$ &    \\
\bottomrule
\end{tabular}
\end{minipage}
\end{center}
\end{table}
\clearpage
\begin{table}[h]
\begin{center}
\begin{minipage}{300pt}
\caption{Implicit Partial Volume Symbols and Constants}\label{Implicit Symbols Table}
\begin{tabular}{@{}lll@{}}
\toprule
Symbol & Description & Default Value\\
\midrule
$\Omega_{t}$  & tissue domain                                           & \\
$\Gamma_{t}$  & tissue domain boundary                                  & \\
$D$           & dimensionality                                          & 3\\
$K$           & gaussian curvature                                      & \\
$\mathcal{W}$ & weingarten map                                          & \\
$\mathcal{P}$ & point cloud                                             & \\
$\mathcal{P}_{S}$& point cloud sample                                   & \\
$\mathcal{S}$ & true manifold                                           & \\
$\textbf{x}$  & coordinate                              & \\
$\lambda$     & thin-plate relaxation                                   & \\
$w$           & semi-compact function                                   & \\
$\psi$        & partition function                                      & \\
$f$           & global implicit function                                & \\
$f_{m}$       & local implicit function                                 & \\
\bottomrule
\end{tabular}
\end{minipage}
\end{center}
\end{table}
\clearpage
\begin{table}[h]
\begin{center}
\begin{minipage}{300pt}
\caption{CFD Symbols and Constants}\label{CFD Symbols Table}
\begin{tabular}{@{}llll@{}}
\toprule
Symbol & Description & Units & Value\\
\midrule
\midrule
$\Omega_{f}$                           & fluid domain                    &                                 & \\
$\Gamma_{f}$                           & fluid domain boundary           &                                 & \\
$\boldsymbol{v}$                       & velocity                        & cm/s                            & \\
$\boldsymbol{f}$                       & body force                      & dyne                            & 0\\
$t$                                    & time                            & s                               & \\
$\rho$                                 & fluid density                   & $\text{g}/\text{cm}^3$          & 1.06\\
$\mu$                                  & dynamic viscosity               & cP                              & 4\\
$\nu$                                  & kinematic viscosity             & $\text{cm} \cdot \text{s}$      & 0.037\\
$P$                                    & pressure                        & Pa                              & \\
$Q$                                    & flow rate                       & $\text{mL}/\text{s}$            & \\
$S$                                    & surface area                    & $\text{cm}^2$                   & \\
$P^0$                                  & ref. config. pressure           & Pa                              & \\
$r^0$                                  & ref. config. radius             & cm                              &\\
$E$                                    & Young's Modulus                 & dyne/$\text{cm}^2$              & \\
$h$                                    & wall thickness                  & cm                              & \\
$k_1$                                  & olufsen parameter 1             &                                 & \\
$k_2$                                  & olufsen parameter 2             &                                 & \\
$k_3$                                  & olufsen parameter 3             &                                 & \\
$z$                                    & axial position                  & cm                              & \\
$R$                                    & hydraulic resistance            & $\text{dyne}/\text{cm}^5$       & \\
$C$                                    & compliance                      & cm$^5$/g/s$^2$                  & \\
$L$                                    & inertia                         & g/cm                            & \\ 
\bottomrule
\end{tabular}
\end{minipage}
\end{center}
\end{table}
\clearpage
\begin{table}[h!]
\begin{center}
\begin{minipage}{450pt}
\caption{Performance of Partial Implicit Tissue Domain Reconstruction}\label{tissue domain reconstruction}
\begin{tabular}{@{}llll@{}}
\toprule
Domain Name & Number of & Build Time & Volumetric \\
& Patches & (sec) & Error (\%) \\
\midrule
\midrule
Branchium of Left Inferior Colliculus  & 1255    & 43.77    & 5.47\\
Third Ventricle                        & 1775    & 53.16    & 0.02\\
Fourth Ventricle                       & 1636    & 50.47    & -0.56\\
Anterior Commissure                    & 1184    & 24.51    & 3.63\\
Commissure of Fornix of Forebrain      & 1327    & 45.48    & -2.12\\
Left Inferior Frontal Gyrus            & 1776    & 55.22    & 2.70\\
Right Inferior Frontal Gyrus           & 1826    & 57.10    & 3.13\\
Left Internal Capsule                  & 1743    & 64.08    & -0.18\\
Right Internal Capsule                 & 1782    & 65.88    & -0.18\\
Left Globus Pallidus                   & 1148    & 40.23    & -0.14\\
Hypothalamus                           & 2316    & 52.26    & 0.57\\
Midbrain                               & 2438    & 53.92    & 0.27\\
Right Stria Medullaris of Thalamus     & 1398    & 34.80    & -3.83\\
Medulla Oblongata                      & 2285    & 65.78    & 0.68\\
Left Superior Frontal Gyrus            & 2038    & 63.98    & 2.87\\
Right Superior Frontal Gyrus           & 1956    & 55.98    & 2.92\\
Right Olfactory Tract                  & 1287    & 23.73    & 1.21\\
Left Olfactory Tract                   & 1401    & 24.71    & 1.71\\
Lamina Terminalis                      & 2227    & 34.92    & 5.54\\
Right Putamen                          & 1762    & 50.13    & 0.00\\
\bottomrule
\end{tabular}
\end{minipage}
\end{center}
\end{table}
\textbf{Table S4. Performance of anatomic tissue domain reconstruction using partial implicit methods.} Twenty non-convex regions of segmented brain tissue were used to build partial implicit models necessary for synthetic vascularization. The number of patches for each tissue domain, reconstruction build time, and volumetric error relative to the original mesh are provided. 
\clearpage
\clearpage
\begin{table}[h]
\begin{center}
\begin{minipage}{400pt}
\caption{Vascularization Performance in Arbitrary Tissue Domains}\label{tissue domain vascularization} 
\begin{tabular}{@{}lll@{}}
\toprule
Domain Name & Build Time (sec) & Benchmark Time (sec) \\
\midrule
\midrule
Branchium of Left Inferior Colliculus  & 56.96    & 3904.24\\
Third Ventricle                        & 87.84    & 3087.96\\
Fourth Ventricle                       & 71.89    & 3211.41\\
Anterior Commissure                    & 61.39    & 2732.70\\
Commissure of Fornix of Forebrain      & 60.00    & 3556.48\\
Left Inferior Frontal Gyrus            & 180.18   & 1367.68\\
Right Inferior Frontal Gyrus           & 270.51   & 1653.80\\
Left Internal Capsule                  & 61.45    & 3298.20\\
Right Internal Capsule                 & 73.00    & 2889.39\\
Left Globus Pallidus                   & 56.52    & 4260.97\\
Hypothalamus                           & 62.00    & 4104.63\\
Midbrain                               & 75.06    & 2868.34\\
Right Stria Medullaris of Thalamus     & 96.82    & 2598.69\\
Medulla Oblongata                      & 76.37    & 3401.14\\
Left Superior Frontal Gyrus            & 284.56   & 1076.14\\
Right Superior Frontal Gyrus           & 188.70   & 1218.69\\
Right Olfactory Tract                  & 57.58    & 3287.84\\
Left Olfactory Tract                   & 80.80    & 3020.03\\
Lamina Terminalis                      & 61.96    & 3775.37\\
Right Putamen                          & 60.43    & 3747.50\\
\bottomrule
\end{tabular}
\end{minipage}
\end{center}
\end{table}
\textbf{Table S5. Synthetic vascular generation performance against public benchmark for single-core, serial CCO algorithm.} Open-loop vascular trees with 1000 terminal vessel segments are constructed within twenty anatomic regions of brain tissue. Time performance of the presented synthetic vascular generation tool is listed under "Build Time" and benchmark times are given under "Benchmark Time" in seconds. 
\clearpage
\clearpage
\noindent \textbf{Materials and Methods}\\
\underline{Iterative vascular generator}\\
A classic implementation of Schreiner’s well-established CCO scheme adapted for 3D volumes is the starting point of the presented work \cite{bib28, bib30}. For clarity, a table of definitions is provided for the symbols used for iterative vascular generation (Supplement Table \ref{Vascular Generation Symbols Table}). Briefly, this approach composes vascular networks as collections of rigid-wall hollow cylinders and assumes homogeneous, incompressible, fully-developed, steady, laminar, axisymmetric blood flow at constant viscosity. Each cylinder represents a vessel segment, $i$, of the total vascular tree, $\mathcal{T}$. Thus, flow everywhere within a vessel network is modeled by analytical solutions to Hagen-Poiseuille flow. Specifically, the resistance, $R_i$, and pressure drop, $\Delta P_i$, of a given vessel can be defined:
\begin{equation}
    R_{i} = \frac{8\mu l_i}{\pi r^4_i}
    \label{resistance}
\end{equation}
\begin{equation}
    \Delta P_i  = R_iQ_i
    \label{circuit_analogy}
\end{equation}
To approximate the hydraulic resistance of the whole vascular tree, a circuit analogy is taken to account for the resistive contributions of vessels in series and parallel. This employs hydraulic analogues of Kickoff's current and voltage laws for flow and pressure, respectively. Thus, along each parent and its set of daughter vessels, $\mathcal{C}$, mass and energy are conserved.  
\begin{equation}
    Q_{parent} = \sum_{i \in \mathcal{C}} Q_i
    \label{conservation_of_mass}
\end{equation}
\begin{equation}
    P_{parent} = P_i + R_iQ_i, \forall i \in \mathcal{C}
    \label{pressure_drop}
\end{equation}
Additionally, the morphometric diameter relationship between a parent and its daughter vessels is assumed follow a power law form proposed by Murray \cite{murray}:
\begin{equation}
    r^{\gamma}_{parent} = \sum_{i \in \mathcal{C}} r^{\gamma}_{i}
    \label{murray}
\end{equation}
Here the exponent, $\gamma$, scales vessel caliber to minimize the work required for mass transport and metabolic upkeep, hence minimizing power dissipation through the network. Physiologically, power law exponents exist in a range $2 \leq \gamma \leq 3$ as vascular networks bifurcate from large arterial vessels towards the finer capillary beds where vessels of higher flow rate generally have lower values of $\gamma$ \cite{murray_exponent}. These assumptions are the foundation of all synthetic vascular generation with CCO methods. Subsequent generation of vascular trees are the realization of random sampling processes constrained to fluid and morphometric scaling laws.

To avoid superfluous calculations during vascular generation, parent-daughter diameter ratios, $\beta$, and reduced resistance, $R^{*}$, are used while appending new vessels. Thus, radii information need only be determined once per iteration.
\begin{equation}
        \beta_i = \frac{r_i}{r_{p}} = \bigg(1+\sum\limits_{j \in \mathcal{C}_i} \big(\frac{r_j}{r_i}\big)^{\gamma} \bigg)^{-1/\gamma}, \forall i \in \mathcal{C}
\label{bifurcation}
\end{equation}
\begin{equation}
    R_{i}^{*} = R_{i}r_{i}^{4}
    \label{reduced_resistance}
\end{equation}
Because (\ref{bifurcation}) is derived from (\ref{murray}), we note that $\sum_{i \in \mathcal{C}}\beta_{i}^{\gamma} \equiv 1$ and $\mathcal{C}_i$ is the set difference between the daughters and current daughter such that $ \mathcal{C}_i = \{ j: j \in \mathcal{C}, \sim (j \in \{ i \}) \}$. When assuming a strictly bifurcating network, $\mathcal{C}$ only contains two vessels and $\mathcal{C}_i$ is merely the other daughter vessel. To calculate vessel radii within the tree we rely on the known pressure drop and flow rate across the entire tree to determine the root radius using (\ref{circuit_analogy}) and (\ref{reduced_resistance}). Equipped with this value, we can propagate radii calculations from the parent-daughter diameter ratios.
\begin{equation}
    r_{root} = \bigg(\dfrac{R_{root}^{*}}{R_{root}}\bigg)^{1/4} = \bigg(\dfrac{Q_{root} R_{root}^{*}}{\Delta P_{tree}}\bigg)^{1/4}
\end{equation}
Similarly, we can derive a general relationship among the diameter ratios of daughter vessels which will be useful when calculating the parent-daughter diameter ratios. Note, if we assume equal pressure drops across daughters, this relationship only depends upon the vessel lengths and flow rates. 
\begin{equation}
    \dfrac{r_j}{r_i} = \bigg(\dfrac{R_{j}^{*}Q_{j}\Delta P_{i}}{R_{i}^{*}Q_{i}\Delta P_{j}}\bigg)^{1/4}, \forall i \in \mathcal{C}
    \label{daughter_ratios}
\end{equation}
For heterogeneous pressure distributions, (\ref{daughter_ratios}) can just as easily be computed so long as the pressure drops are obtained from a known distribution profile. Combining (\ref{resistance}), (\ref{circuit_analogy}), (\ref{pressure_drop}), (\ref{bifurcation}), and (\ref{reduced_resistance}) we can obtain a general relationship for the reduced resistance of any vessel in the entire vascular tree, $\mathcal{T}$. 
\begin{equation}
    R_{i}^{*} = \dfrac{8\mu L_{i}}{\pi} + \bigg(\sum\limits_{k \in \mathcal{C}} \dfrac{\beta_k}{R_{k}^{*}}\bigg)^{-1}, \forall i \in \mathcal{T}
\end{equation}
Note that terminal vessels within a tree do not have any children; thus, the set of of daughters is empty, $\mathcal{C} = \emptyset$, and the reduced resistance is exactly (\ref{reduced_resistance}) which only depends on vessel length.

\noindent \underline{Implicit Partial Volume}\newline
In addition to fluid and scaling constraints, a tissue domain must be supplied. This properly bounds the optimization problem within a feasible region, and it can practically be thought of as the tissue domain within which vascular networks are being generated. In literature, synthetic vasculature implementations rely on varying assumptions about the tissue geometry. Most notable are assumptions about concavity and hollowness \cite{bib30, bib32}. The most liberal assumptions define perfusion volumes as convex solids with uniformly distributed vessels. In the body, however, we observe that organs are not necessarily convex (e.g. reticulated topology of the brain and "bean" shape of the kidneys). Furthermore, organs such as the heart and lungs have fluid cavities for blood and air, respectively. Previous work has been made to recapitulate nonconvex surfaces with and without cavities using random walkers, Monte Carlo, and Poisson (signed-distance) methods. In early work on hollow organs, ideas from electrostatics were borrowed to reconstruct implicit surfaces; however such implementations required expensive triangulation of known surface models. In this work we leverage recent techniques in implicit surface reconstruction methods to produce energy field representations of complex perfusion volumes given point-cloud data. Evaluation of the resulting implicit functions yields efficient boundary detection necessary for synthetic vascularization. Level sets of these fields yield contours which adequately recapitulate 3D surface models (S Figure 5).  

We begin with a tissue domain of interest contained within an associated boundary manifold $\mathcal{S} \in \mathbb{R}^{D}$ obtained from medical or engineering data. $\mathcal{S}$ is the true manifold and is typically unknown for medical data which cannot be directly described by analytic functions. Instead, quantitative measurements provide point clouds $\mathcal{P}$ approximating the true manifold. Taking a sufficient set of the total points $\mathcal{P}_S = \{\textbf{x}_{1},...\textbf{x}_{N}\} \subseteq \mathcal{P}$, we wish to reconstruct a smooth analytical function, $f(\textbf{x}) \in \mathbb{R}^{d}$, for which the zero-level set closely approximates the true bounding surface, $L_{0}(f) \approx \mathcal{S}$, of the perfusion volume. Additionally, we want to formulate the reconstruction procedure without selection of a shape parameter, a common requirement of radial basis function (RBF) approaches.

To obtain $f$ we formulate a variational polyharmonic spline partition of unity method. Using the subset, $\mathcal{P}_S$, we can approximate the local surface at $\textbf{x}_{i}$ with a polyharmonic spline (PHS), $\phi_l(r)$, which is given in the following general forms \cite{polyharmonicSplines}:

\begin{align}
    \phi_l(r) = (-1)^{l+1} \begin{cases}
                             r^{2l}\log{r}, & \text{$l$ even positive} \\
                             r^{2l+1}, & \text{$l$ odd non-negative}
                         \end{cases}
\end{align}
\newline
Here $r=||\textbf{x} - \textbf{x}_i||$ is the $L_2$-norm about the interpolating point such that we consider the PHS as being centered at $\textbf{x}_i$. Order parameter, $l$, is based on the assumed smoothness of the data, though, in application, it is typical to see $l$ of low order as higher order interpolants tend to suffer from depreciating numerical stability \cite{numericalStability1,numericalStability2}. To avoid the trivial solution $f \equiv 0$, additional constraints need to be made as in previous variational implicit surface reconstruction literature \cite{constraint1,vpss}. The general variational definition is given as the minimized sum of the interpolating surface function, $f$, and the approximated energy associated with that function, $\lambda E(f)$. We also constrain that the gradient of $f$ to have unit magnitude. 
\begin{equation}
\begin{aligned}
    \text{Minimize:} \sum\limits_{i \in \mathcal{P}_S} f(\textbf{x}_i)^2+\lambda E(f)\\
    \text{Constraint:} ||\nabla f(\textbf{x}_i)|| = 1,   \forall \textbf{x}_i \in \mathcal{P}_S
\end{aligned}
\label{minimize}
\end{equation}
Thus we have outlined the general optimization problem for variational surface reconstruction from a given point set. For our interpolating function, $f$, we select Duchon's interpolant, $f_{du}$ which is composed of a linear combination of PHS of degree $l=1$ and lower degree terms \cite{duchon}.
\begin{equation}
\begin{aligned}
    f_{du}(\textbf{x}) = &\sum_i^N a_i \phi_1(||\textbf{x}-\textbf{x}_i||)+\\
    &\sum_i^N\textbf{b}_i^T\dfrac{\partial}{\partial\textbf{x}_i}\phi_1(||\textbf{x}-\textbf{x}_i||)+\textbf{c}^T\textbf{x}+d\\
    \text{Subject to:}\\
                           &\sum_i a_i = 0\\
                           &\sum_i a_i\textbf{x}_i + \sum_i \textbf{b}_i = 0
\end{aligned}
\label{duchon_interpolant}
\end{equation}
Duchon's interpolant guarantees an energy of at least 2$^{\text{nd}}$-order without need for domain discritization, ensuring computability. Additional orthogonality constraints restrict the problem to a unique solution. Both $f_{du}$ and $||\nabla f_{du}||$ can be expressed in closed-form allowing (\ref{minimize}) to become a constrained quadratic optimization problem. The energy function, $E$, is the differential energy for which Duchon's interpolant is known to be optimal. In one dimension, this differential energy is the classical "thin plate" energy. For functions of higher dimensions, this energy is analogous on the Fourier transform of the interpolant function. 
\begin{equation}
    E(f) =\bigg(\int_{\mathbb{R}^D} \vert \boldsymbol\tau \vert ^{2s} \bigg\vert
    \int_{\mathbb{R}^D}\dfrac{\partial^2}{\partial^2\textbf{$\boldsymbol{\xi}$}}f(\textbf{$\boldsymbol{\xi}$})e^{-2\pi j\boldsymbol\tau\boldsymbol\xi}d\boldsymbol\xi\bigg\vert^2
    d\boldsymbol\tau\bigg)^{1/2}
    \label{multidimensional_duchon_energy}
\end{equation}
In all cases, the energy is exactly the semi-norm of the function space, as was originally stated by Duchon \cite{duchon}. To more easily solve this system we can represent the interpolant and orthogonality constraints (\ref{duchon_interpolant}) as a system of linear equations.
\begin{equation}
    A
    \begin{pmatrix}
    \textbf{a}\\
    \textbf{b}\\
    \textbf{c}\\
    \text{d}
    \end{pmatrix}
    =
    \begin{pmatrix}
    f_{du}(\textbf{x}_i)\\
    \nabla f_{du}(\textbf{x}_i)\\
    \textbf{0}\\
    0
    \end{pmatrix}
    \label{linear_relation}
\end{equation}
Here $A$ is the coefficient matrix of the interpolant. For clarity, the general calculation for the elements of $A$ have been presented in (S Figure A). From previous work in variational approaches, we recognize the specific case of (\ref{minimize}) where $f$ and $E(f)$ are Duchon's interpolant and energy, respectively. For greater discussion on interpolant selection and formulation of variational implicit point set surfaces we direct the reader to original global derivation presented below \cite{vpss}.  
\begin{equation}
\begin{aligned}
    \text{Minimize:  } & \nabla f(\textbf{x}_i)^T \cdot (M_{w \times w}-\lambda M_{o \times w}^T\\
    &(I-\lambda M_{o \times o})^{-1}M_{o \times w}) \cdot \nabla f(\textbf{x}_i)\\
    \text{Constraint:  } & ||\nabla f(\textbf{x}_i)|| = 1,   \forall \textbf{x}_i \in \mathcal{P}_S\\
    \text{Given:  } & o \leq N \\
                  & N \leq w \leq N + N\times D 
\end{aligned}
\label{minimize_duchon}
\end{equation}
Here we denote the cost function being minimized as $B$. $M$ is the inverse of the coefficient matrix, $A$; $M_{i\times j}$ is the submatrix spanned among the elements specified by the ranges of $i$ and $j$. Because (\ref{multidimensional_duchon_energy}) is equipped with a semi-inner product and linearly related to $f$ and $\nabla f$ through (\ref{linear_relation}), a variational optimization problem is posed in (\ref{minimize_duchon}). This can be minimized for a given value of $\lambda$, which we will refer to as the thin-plate relaxation factor (TRF). To arrive at the optimal interpolant for a given point cloud, it is necessary to determine which TRF value yields minimal energy. In previous work, authors selected a few values in the closed interval [0,1] and selected the TRF corresponding to the minimal value within the limited trial set. However, Duchon's energy is sensitive to the TRF chosen and its resulting function space is nonconvex. Hence, picking a handful of values is likely to overlook multiple minima for (\ref{minimize_duchon}); a more rigorous strategy is required. Specifically, we are interested in finding the value $\lambda^{*}$ such that $B(\lambda^{*}) \leq B(\lambda), \forall \lambda \in [0,1].$ We define this as the meta-optimization of (\ref{minimize_duchon}). Thus, we have set up a rigorous approach to globally minimize and reconstruct the approximate surface to $\mathcal{S}$ from point set $\mathcal{P}_S$.

Unfortunately, computational complexity limits implicit reconstruction to less than a few thousand points for global approaches. This bottleneck is due to the inversion of $A$ which scales in time and memory at $\mathcal{O}(D^3N^3)$. Furthermore, as point clouds increase in size the coefficient matrix becomes ill-conditioned \cite{illcondition}. To overcome these, we implement this variational approach as a partition of unity (PU). Traditionally PU methods decompose a given domain, $\Omega$, into small overlapping subdomains $\Omega_m$, for $M \ll N$ where $\bigcup_{m=1}^M \Omega_m \supseteq \Omega$ for $\{m \in \mathbb{Z}: 1 \leq m \leq M\}$. Local approximations can then be made within subdomains and blended together to yield a final reconstructed surface \cite{partitionUnity}. Instead of domain descritization, we directly take patches from the point cloud using estimations of curvature and K-nearest neighbor (kNN) techniques such that $\bigcup_{m=1}^M \mathcal{P}_m \supseteq \mathcal{P}_S$. To efficiently estimate curvature we approximate the Weingarten map, $\mathcal{W}$, using local principal component analysis (PCA) and least squares minimization (LSM) on $\mathcal{P}_S$ \cite{weingarten}. Because LSM involves $N$ gradient approximations for the residual, $\mathcal{W}(\mathcal{P}_S)$ estimates can be obtained in an efficient manner via a closed-form solution. We note the Guassian curvature, $K_i$, at a given point has a defined relationship to the determinant of the Weingarten map at that point.
\begin{equation}
    K_i = \text{det}(\mathcal{W}_i), \forall i \in N
    \label{curvature}
\end{equation}
We can thus decompose the global point cloud into more manageable patches using curvature estimations obtained from (\ref{curvature}) and a point cloud patch decomposition routine. To maintain efficiency we limit the upper number of kNN queries and supply a curvature threshold, $\epsilon$, for patch inclusion determined from the span of Gaussian curvature calculated for the entire point cloud.
With defined patches, we can solve for local surface approximations using the formulated variational minimization (\ref{minimize_duchon}) and interpolant (\ref{duchon_interpolant}) to yield a set of local solutions $f_{m}$.
For each solution there is an associated weighting function, $w_m$, which serves as a suitable approximation of a Delta function about the center of a given patch such that $f(\textbf{x}) \approx \int_{\mathbb{R}^D}f(\boldsymbol{\xi})w_{\textbf{x}}(\boldsymbol{\xi})\partial\boldsymbol{\xi}$. We consider a family of Shepard-like weighting functions to allow for varying assumptions in surface noise and support compactness \cite{shepard,shepard2,shepard3,shepard4}.
\begin{equation}
\begin{aligned}
    \centering 
    w_m = \begin{cases}
        \begin{aligned}
            &\dfrac{1}{(||\textbf{x}-\textbf{x}_m||)^u} \\[.5em]
            &\bigg(\dfrac{1}{||\textbf{x}-\textbf{x}_m||}-\dfrac{1}{R_m}\bigg)^u_+ \\[.5em]
            &\dfrac{1}{D_{RD}^{(p)}(\textbf{x},\textbf{x}_m)^{u}}
        \end{aligned}
        \end{cases}
        \label{weighting_functions}
\end{aligned}
\end{equation}
The set of weighting functions allow us to blend the local approximates together using Shepard's method to generate partition functions, $\psi_{m}(\textbf{x})$, constrained to unity, $\sum_{m=1}^{M} \psi_m \equiv 1$ \cite{shepard2}. The partition function for a patch is given:
\begin{equation}
    \psi_{m}(\textbf{x}) = \dfrac{w_{m}(\textbf{x})}{\sum\limits_{i=1}^{M} w_{i}(\textbf{x})}
\end{equation}
Finally the total reconstructed implicit volume can be calculated as the linear combination of local approximates and associated partition functions.
\begin{equation}
    f = \sum_{m=1}^{M}f_{m}(\textbf{x})\psi_{m}(\textbf{x})
    \label{final_eq}
\end{equation}
Depending on the choice of weighting in (\ref{weighting_functions}) we are guaranteed that (\ref{final_eq}) is at least $C^{2}$ differentiable which we exploit in later vascularization procedures. Because $f$ is a combination of local solutions we can efficiently determine the interior and exterior of the perfusion domain, a common calculation during the synthetic vascular generation process which can be further accelerated by additional Kd-tree heuristics.\\
\underline{Generating close-loop synthetic vascular networks} \\
Close-loop networks are realized by forming connecting vessels between generated synthetic vascular trees within the same tissue domain. In general, this problem is expensive to solve and its complexity intimately depends on the geometric complexity of the tissue domain as well as the definition of the cost function for creating vessel connections among terminals. We proceed to form connecting vessels in two stages: 1) terminal assignment and 2) geometric realization. During terminal assignment, one terminal from each tree is allocated into a set to form $N$ sets; the goal is to reach a total minimum cost among the combinations of $N$ terminal sets for the $D$ interpenetrating trees. This is generally considered a nonlinear multidimensional assignment problem (MAP) for which extensive literature exists \cite{MAP,MAP_book}. The problem can be stated as follows: 
\begin{equation}
    \sum_{a \in A} \mathcal{C}(a,\pi_{0}(a),\pi_{1}(a),...,\pi_{D-1}(a))
\end{equation}
 We assume that the cost function $\mathcal{C}$ of forming connecting vessels seeks to minimize total path-length for all permutations of terminals $\pi_{D}$. Because MAPs are NP-hard, we seek to reduce the complexity of assignment to a linear assignment problem (LAP) for which known polynomial time solutions exist. For convex tissue domains, this is relatively straightforward as the cost function is exactly the euclidean distance matrix; thus, no other weighting functions need to be considered. We use a Hungarian algorithm with time complexity $\mathcal{O}(N^3)$ to solve LAPs \cite{scipy}. For nonconvex domains, we construct a geodesic distance matrix. While much more costly, this approach allows the nonconvex problem to satisfy the LAP conditions and avoids quadratic assignment complexities otherwise introduced through additional penalty functions \\
 Next, geometric realization connects the set of terminals with connection vessels that satisfy a list of competing criteria including collision avoidance, minimum path lengths, and tortuosity. We solve the geometric realization problem by fitting Bezier curves through control points between the assigned terminals \cite{geomdl}. Control point coordinates define the set of variables for a global multiobjective optimization procedure which can be solved using a simplical homology or basin hopping approaches depending on the number of control points \cite{scipy}.\\
\underline{Automatic multifidelity hemodynamic model creation}\\
The presented synthetic vascularization toolkit is built towards specific integration with the open source project SimVascular (www.simvascular.org) which provides a sophisticated CFD package for simulating blood flow in physiologic models \cite{updegrove}. Synthetic vascular networks are created within a SimVascular pipeline framework through an integrated Python interpreter which calls C++ bound functions underlying much of the core source code. We briefly cover the main steps in this pipeline and discuss differences between this automated procedure versus manual model construction.\\
\textit{Pathline generation} \\
Unlike the traditional pipeline which begins with image processing, we already have a vascular network represented as a collection of discrete cylinders which does not rely on image-based data. Therefore, we proceed immediately to pathline creation (Figure 3A i). During this step we group cylinders into vessel segments and generate smooth, pathline interpolations along the center of cylinder groups. In the manual pipeline, pathlines provide a reference frame to resampling image data onto cross-sections to delineate luminal boundaries for vessels. For generated synthetic vasculature, this is unnecessary. We, instead, assume that interpolated pathlines are sufficient to guide contour segmentation without need for manual verification. Pathline interpolations are constructed using B-spline representations of general N-D curves\cite{ndcurves}.\\
\textit{Contour setting}\\
Along generated pathlines, contours are evenly placed and proportional to the number of knots defining a given pathline. This approach provides greater cross-sectional resolution to vessels of higher curvature and ensures accurate lofting. Each 2D segmentation is represented using analytic circular geometries with radii obtained from synthetic vasculature data (Figure 3A ii). When segmenting patient-specific data, other types of cross-sections including level set contours or closed-curve splines have to be employed to capture the lumen geometry of vessels. However, synthetic vascular generation assumptions idealize vessel cross-sections as circles; thus, circular contours are sufficient. To verify the integrity of contours, an overlap check is performed within each vessel to remove any pairs of contours which would result in lofting errors (Figure 3A iii). These down-sampled regions are then resampled with contours optimally placed to avoid overlaps and maintain vessel continuity.\\
\textit{Solid generation}\\
After contour segmentation and overlap checking, splines are formed along the length of each vessel connecting the 2D segmentations and creating a Non-Uniform Rational B-Splines (NURBS) surface (Figure 3A iv). These analytic surfaces are then triangulated and capped to form complete, watertight vessels needed for CFD using PolyData solid modeling kernels. To accomplish this, custom scripting combined with geometry filters available through VTK (www.vtk.org) and VMTK (www.vmtk.org) have been included as modules within the SimVascular Python environment. The set of individual solid vessels are iteratively combined through ordered Boolean addition (Figure 3A v). Because unions of discrete surface data can results in sharp, un-physiologic geometries at bifurcations, localized smoothing is performed at each bifurcation with either a constrained or unconstrained Laplacian smoothing filter (Figure 3A vi). Additional localized operations such as subdivision, hole filling, and decimation are available depending on model complexity. At the end of model creation, faces of the vascular model (namely caps and walls) are automatically assigned names. This labeling methodology allows boundary conditions to be easily applied prior to simulation. Although all of these steps are automated, it is recommended to visually inspect models before beginning continuing computationally expensive steps.\\
\textit{Meshing}\\
Fluid simulation relying on finite element methods require discretization of the fluid domain through mesh generation. While SimVascular supports both commercial and open-source meshing tools, this automated pipeline relies exclusively on the open-source meshing kernels from Tetgen (www.tetgen.org) as well as boundary layer meshing from VMTK and remeshing optimization routines from MMG (www.mmgtools.org). Meshing is a consequential step for solution accuracy. Therefore, great care should be taken when discretizing models as one set of parameters are not guaranteed to perform well for synthetic vascular realizations across different perfusion domains or different boundary assumptions. By default, meshes will be generated with three boundary layers near wall regions. These thin layers of small elements are employed to better characterize the normal velocity gradient close to the walls. Still determining the number of layers and the size of the elements within these layers is non-trivial and may require further user verification to ensure convergence of velocity profiles near wall boundaries.
\textbf{Fluid Simulations} \\
\textit{3D CFD Formulation} \\
Fluid flow within 3D vascular networks is modeled using the incompressible Navier-Stokes equations shown here in vector notation,
\begin{align}
    \rho \bigg( \dfrac{\partial \boldsymbol{v}}{\partial t} + \boldsymbol{v} \cdot \nabla \boldsymbol{v}\bigg) &= -\nabla P + \mu \nabla^2\boldsymbol{v} + \boldsymbol{f} 
    \label{momentum}\\
    \nabla \cdot \boldsymbol{v} &= 0
    \label{continuity}
\end{align}
where $\rho$ and $\mu$ are the density and dynamic viscosity of the fluid ($\rho = 1.06 \frac{g}{cm^3}, \mu=4 \text{ cP} $ for blood, $\rho = 1.0 \frac{g}{cm^2}, \mu=0.89 \text{ cP}$ for culture media or water). Equation (\ref{momentum}) defines the conservation of momentum with $\boldsymbol{v}$, $\boldsymbol{f}$, $P$, and $\nabla^2 \boldsymbol{v}$ representing the time derivative, body forces, pressure, and viscous portion the stress tensor, respectively. The continuity equation (\ref{continuity}) is derived from the conservation of mass. To solve the set of partial differential equations governing this system, our open-source solver, svSolver, relies on a finite element formulation for the fluid domain (www.//github.com/SimVascular/svSolver). svSolver uses linear tetrahedral $P_1$-$P_1$ elements with streamline-upwind/Petrov-Galerkin (SUPG) and pressure-stabilizing/Petrov-Galerkin (PSPG) methods \cite{supg_pspg}. This stabilized formulation is progressed in time using a generalized alpha time stepping scheme \cite{alpha}. More information on the finite element formulation underlying svSolver is detailed in references\cite{updegrove,supg_pspg,precondition}.\\
\textit{1D CFD Formulation}\\
Depending on the spatial resolution required of the solution, one-dimensional (1D) simulations may be sufficient. These models provide only the axial dimension along centerlines of vessels but are capable of resolving pressure and flow wave propogations down vascular trees. By integrating the 3D Navier-Stokes equations over the cross-section, we can represent the fluid problem through bulk quantities including volumetric flow rate $Q$, average cross-section pressure $P$, and cross-sectional area $S$.
\begin{align}
    \dfrac{\partial Q}{\partial t} + \dfrac{4}{3} \dfrac{\partial}{\partial z} \dfrac{Q^2}{S} + \dfrac{S}{\rho} \dfrac{\partial P}{\partial z} &= Sf-8\pi \nu \dfrac{Q}{S} + \nu \dfrac{\partial^2 Q}{\partial z^2} \\
    \dfrac{\partial S}{\partial t} + \dfrac{\partial Q}{\partial z} &= 0
\end{align}
In order to close this 1D formulation, we need to apply a constitutive relation between the pressure and cross-sectional area. From the multiple constitutive models commonly used in literature, we default synthetic vascular networks to an Olufsen material model shown here \cite{olufsen}.
\begin{equation}
\begin{split}
    P(z,t) &= P^0(z)+\dfrac{4}{3}\dfrac{Eh}{r^0(z)}\Bigg( 1 - \sqrt{\dfrac{S^0(z)}{S(z,t)}} \Bigg), \\ \dfrac{Eh}{r^0(z)} &= k_1e^{k_2r^0(z)}+k_3    
\end{split}
\end{equation}In this material model, $E, h, r^0, S^0$ are the Young's modulus, wall thickness, reference radius, and reference cross-sectional area, respectively, at a given reference pressure, $P^0$. The constants $k_1, k_2,$, and $k_3$ are empirically derived; however, for rigid wall simulations $k_1$ is set to zero and $k_3$ is a sufficiently high, but physically arbitrary, value to numerically impose rigid behavior for the vessel walls. Centerline extraction for large networks is accomplished through custom scripting in SimVascular and VMTK. This system of partial differential equations is solved using our open-source 1D solver, svOneDSolver (https://github.com/SimVascular/svOneDSolver) \cite{oneDmodel, reduced_models}.\\
\textit{0D CFD Formulation} \\
Often to understand bulk flow behavior within these networks, we rely on zero-dimensional (0D) simulations. Like the assumptions used to construct initial synthetic vascular networks, these 0D analogs to the Navier-Stokes equations lack spatial resolution and can only represent average pressure drops and flow quantities over vessel segments. As such, these simulations are cheap to execute making them ideal for networks with hundreds or thousands of vessels. Three common cardiovascular properties captured by these models are hydraulic resistance, vessel compliance, and blood inertia. The flow rates and pressures within vessels are governed by the following linear differential and algebraic equations. 
\begin{align}
    \Delta P  &= QR \\
    Q &= C\Delta \Dot{P}\\
    \Delta P &= L\Dot{Q}\\
    Q &= \dfrac{|Q|+Q}{2}
\end{align}
Here $R$ is the hydraulic resistance, $C$ is the vessel compliance, and $L$ is the inertial contribution of blood. Combinations of resistive, capacitive, and inductive elements form lumped parameter models which may describe the bulk behavior of individual vessels or entire vascular trees. In 0D simulations of synthetic vascular networks, individual vessels are modeled with dedicated lumped parameter elements to ensure that bulk flow quantities can be obtained for all vessels. The set of governing 0D equations are solved using our in-house numerical solver svZeroDSolver (https://github.com/SimVascular/svZeroDSolver). Unlike the assumptions governing synthetic vasculature generation, 0D simulations allow for inertial and elastic properties of blood and blood vessel walls to be captured. Thus waveform propagation can be simulated for bulk quantities in large vascular systems necessary in large-scale tissue engineered applications\\
\textbf{3D Bioprinting}\\
\textit{FRESH Preparation}\\
FRESH support bath was generated using a complex coacervation method as previously described \cite{hudson}. The FRESH support bath was made by dissolving 3.0\% (w/v) gelatin type B (Fisher Scientific, G7-500), 0.3\% (w/v) gum arabic (Sigma Aldrich, G9752), and 0.125\% (w/v) Pluronic® F-127 (Sigma Aldrich, P2443) in a 50\% (v/v) ethanol solution at 45oC in a 1 L beaker. The pH of the solution was adjusted to 5.65 using 1M hydrochloric acid (HCl) and stirred overnight at room temperature. The next day, the slurry was centrifuged at 300 g for 2 min, the ethanol supernatant discarded, and the support bath was washed 3 times with deionized (DI) H2O. After the third round of washing, the supernatant was replaced with a solution of 50 mM HEPES (Corning, 60-034-RO) buffered to a pH of 7.4.\\
For printing, the uncompacted support bath was degassed at room temperature for 30 min followed by centrifugation at 2000 g for 5 min to compact the particles. The supernatant was discarded, and the slurry was transferred into the print container of choice. A bioink of 23 mg/mL acidified collagen type I was utilized for printing by diluting a 35 mg/mL acidified collagen bioink (LifeInk 240, Advanced Biomatrix, 5267) in a 2:1 volume ratio with sterile DI H2O. FRESH printing was executed on custom-built 3D bioprinter (Aerotech) using a 30 gauge 150 µm ID needle, (Jensen Global, JG30-0.5HPX) mounted to a 2.5 mL glass gastight syringe (Hamilton, 81401).\\
\textit{Planar Perfusion} \\
For perfusion, a peristaltic pump (Ismatec, EW-95663-34) with 1.42 mm ID peristaltic tubing (Cole Parmer, EW-95663-34) and 1/16” ID silicone tubing (Cole Parmer, EW-95802-02) was connected to the bioreactor inlet while silicone tubing connected to the bioreactor outlet returned to the pump reservoir. Black dye (McCormick, 052100581873) was perfused at 250 µL/min. The simulated vascular network tissue FRESH printed from collagen type I was inserted into a custom bioreactor and perfused as described in previous work \cite{hudson}. \\
\textit{3D print code generation for FRESH}\\
The simulated vascular network model was subtracted from a solid tissue block designed in CAD software (Autodesk Fusion 360) to allow for the vessel network to be inserted and perfused in the perfusion bioreactor system. All vascular network STL models were processed in slicing software (Ultimaker, Cura) using a layer height of 60 µm, a print speed of 15-50 mm/s, 3 perimeters, 3 top and bottom layers, 35\% infill density, 95\% extrusion multiplier (AKA flow tweak), and 0.1 mm retraction length. The exported G-code was copied into an Aerotech basic program editor for execution on the bioprinter.\\
\textit{OCT Imaging and 3D Gauging}\\
OCT imaging was performed as previously described in Tashman et al 2023 \cite{oct_guaging}. Briefly, 3D images were acquired using a Thorlabs Vega 1300 nm OCT system (VEG210C1) with a OCT-LK4 objective. The OCT settings such as amplification and reference intensity were set to provide the highest signal without introducing image artifacts. Upon image acquisition, the 3D image was exported as a 32-bit TIFF stack for further processing and segmentation. The TIFF stack is first imported into FIJI ImageJ for background subtraction and denoise processing. The pre-processed image is then imported into 3D Slicer (www.slicer.org/) for internal network segmentation using the grow from seeds built-in segmentation tools. After segmentation was completed, the full 3D model of the internal network was exported as a stereolithography (STL) file. The segmented STL file from OCT imaging and the originally designed CAD STL file were then imported into CloudCompare (www.cloudcompare.org/) for 3D gauging. Both models were aligned to the origin and registered to each other while maintaining their scale properties. Distance to distance deviations were then calculated and displayed as a color map and histogram showing the positive and negative deviations of the printed model from the original intended CAD model. \\
\textit{Brightfield and Darkfield FRESH Imaging}\\
Brightfield and darkfield images of the FRESH printed generative vascular network were acquired using a Leica M165 FC stereomicroscope, 1.0X objective, with a Prime 95B camera (Photometrics) run by MicroManager 2.0 software. TIFF images were saved and exported for further analysis and image display. \\
\textit{Granular hydrogel preparation}\\
The support matrix and ink both consisted of Lubrizol Carbopol\textsuperscript{\tiny\textregistered} 974P NF Polymer (\#CBP1053H) prepared at 0.2 wt\% concentrations. The Carbopol was prepared by dissolving 2g of Carbopol powder and 0.8g NaOH in 100 mL of deionized water and mixing at 1500-2000 rpm in the FlackTek SpeedMixer\textsuperscript{\tiny\textregistered} (DAC 330-100 Pro) for two minutes before diluting 10X with deionized water. The solution was then stirred overnight at room temperature with a Fisherbrand\textsuperscript{\tiny\texttrademark} CompactDigital\textsuperscript{\tiny\texttrademark} Overhead Stirrer  (Fisher Scientific \#14500211) at 685-950 rpm. To enable visualization of the printed vascular network, colorful acrylic powdered pigment was added to the Carbopol used for the ink formulation. Specifically, the Magenta color of the Pearl Ex Powdered Pigments (\#632) was added to the Carbopol at 2 wt\%. The clear Carbopol used for the support matrix was placed into a transparent acrylic box, while the pigmented Carbopol used for the ink was loaded into a 5 mL glass syringe (Hamilton \#81520, Model 1005 TLL) for dispensing into the support matrix. The nozzle attached to the syringe was a 30-gauge Nordson\textsuperscript{\tiny\textregistered} EFD\textsuperscript{\tiny\textregistered} Optimum\textsuperscript{\tiny\textregistered} General Purpose Stainless Steel Tip (\#7018433), which has a length of 0.5” and an inner diameter of 0.15 mm.\\
\textit{3D print code generation for granular hydrogel printing}\\
To produce the print path, 3D coordinates defining the vascular network were exported from SimVascular and converted to G-code using a custom Python script. The G-code was further converted to the Aerobasic programming language, which interfaces with the 3D printer through the Aerotech A3200 Motion Composer software program. The printer is a custom three-axis motion controller system with four individually addressable printheads (manufactured by Aerotech, Inc.). The syringe containing the ink was mounted to one of the printheads and dispensed at a nozzle translation speed of 2.5 mm/s and filament extrusion speed of 0.0118 mm/s. After printing, the network was imaged using a Canon EOS 5D Mark IV Body with a Canon EF 180 mm f/3.5L Macro USM lens. \\
\textit{microCT of granular hydrogel printing}\\
Quality of the hydrogel prints was assessed using micro-computed tomography (micro-CT) imaging on a Bruker Skyscan 1276. Slice images were acquired with a resolution of 20.54 $\mu$m/pixel, a rotation step of 0.4 degrees, and averaging of 2 frames using a 1.0 mm Al filter with the source voltage set to 85 kV and source current set to 180 $\mu$A. DICOM images were reconstructed from a cropped region of interest in Bruker’s NRecon software package with ring artifact correction set to 10.
\\
\textbf{Synthetic Vasculature Performance Testing}\\
All performance testing of synthetic vascular code was completed on Windows 10 Dell Precision 7920 Tower with Intel\textsuperscript{\tiny\textregistered} Xeon\textsuperscript{\tiny\textregistered} Gold 5220 CPU @ 2.2GHz, 2195 Mhz Clock Speed. A standard Constrained Constructive Optimization implementation benchmark is taken from a benchmark used in Cury \textit{et al} 2021 to compare serial performance of the presented work against previous implementations \cite{pdcco,dcco}. Since the presented work only offers serial generation the selected benchmark implementation offers an ideal one-to-one comparison.  

\clearpage
\begin{figure}[H]
    \centering
    \includegraphics[width=0.8\textwidth]{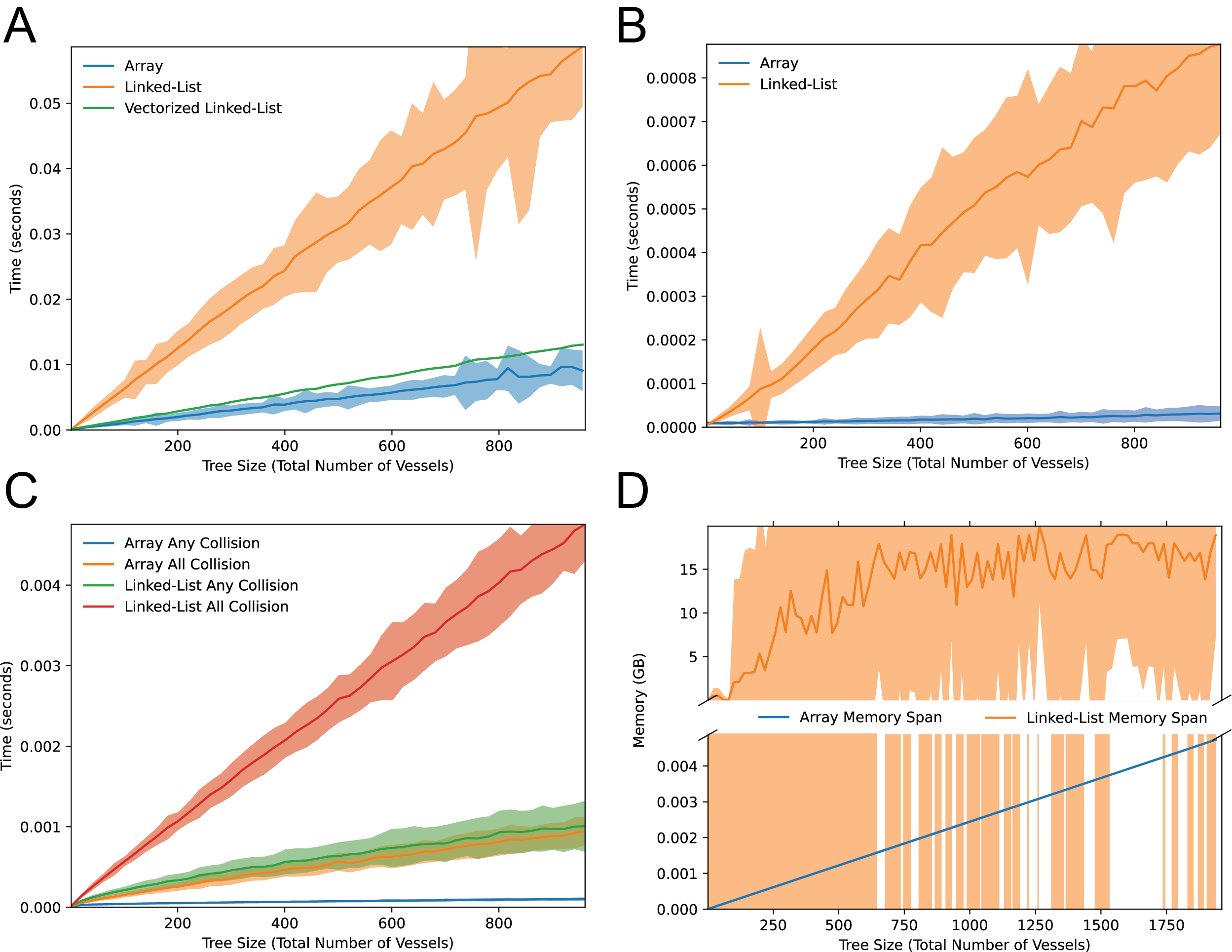}
    \label{S Fig. 1}
\end{figure}
\textbf{Fig. S1. Continuous array versus linked-list container performance. A.} Determining optimal branch appendage for new vessels between presented multi-modal array containers (blue) versus traditional linked-list (orange) and hardware accelerated (vectorized; green) linked-list searches. Shaded regions represent $\pm$ 2SD. Each test is reran 100 times. All algorithms display $\mathcal{O}(n)$ scaling complexity \textbf{B.} vascular tree radii rescaling performance between multi-modal array and linked-list containers. Due to recursion limitations, hardware acceleration is not available for linked-list containers during vascular rescaling. Each test is reran 100 times. Shaded regions represent $\pm$2SD. \textbf{C.} Collision detection between multi-modal array and linked-list containers. Detecting "Any" collisions are generally cheaper heuristic searches which exit upon finding the first collision and at worse scales at $\mathcal{O}(n)$. Detecting "All" collisions is more expensive and confers consistent linear scaling complexity regardless of the number of collisions present. Each test is reran 100 times and shaded regions represent $\pm$ 2SD \textbf{D.} Memory locality of multi-modal versus linked-list containers. Memory size is calculated from the furthest memory addresses present in the container. Because arrays are defined continuously, we expect and observe a linear increase in memory locality with no variance in storage size. Linked-lists display significant storage variability. Shaded regions represent $\pm$ 2SD. 
\clearpage
\begin{figure}[H]
    \centering
    \includegraphics[width=0.7\textwidth]{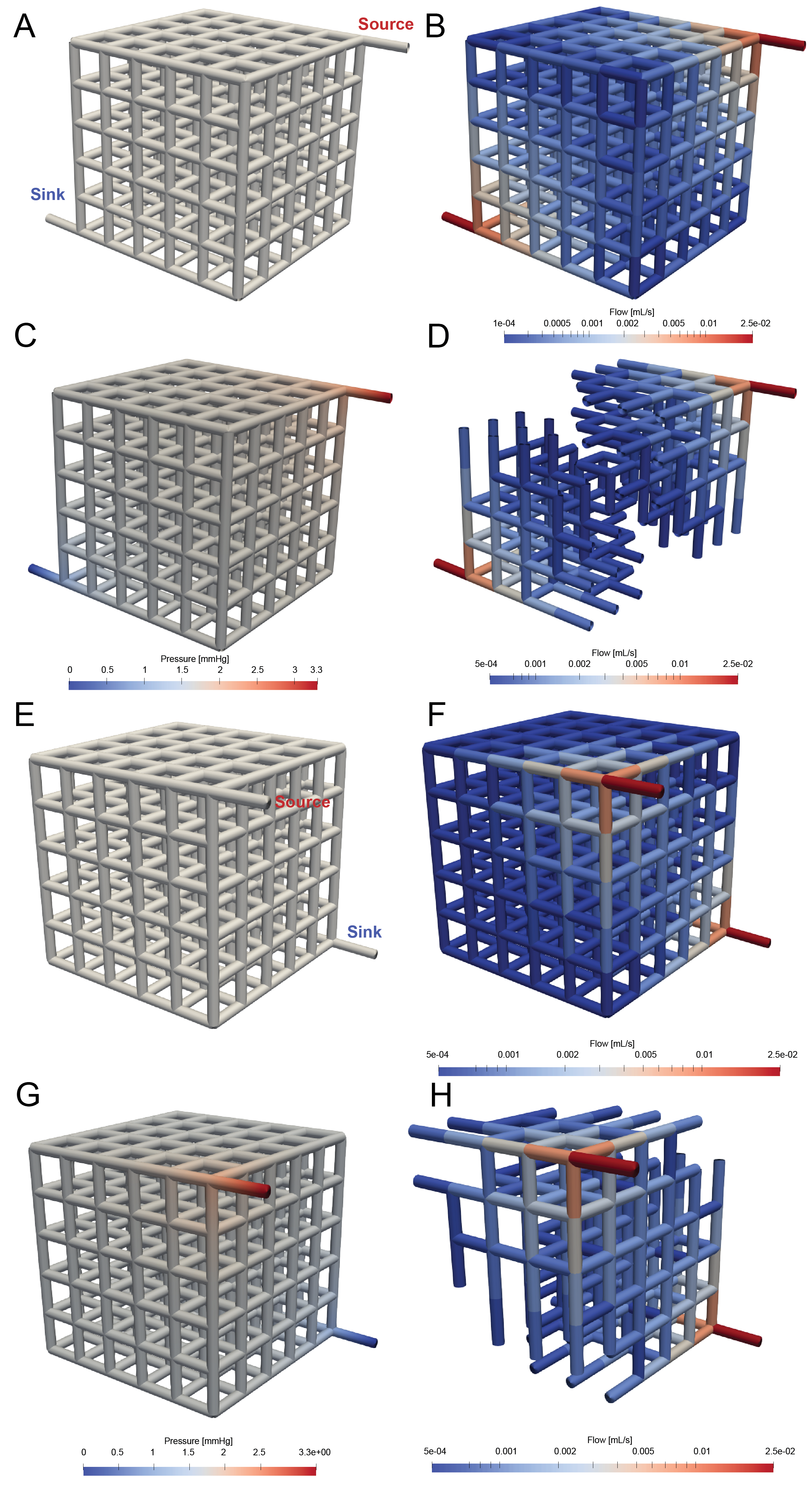}
    \label{S Fig. 2}
\end{figure}
\textbf{Fig. S2. 0D simulation pressure and flow distributions in contrived lattice architectures. A} simple lattice architecture with 125 vessels for a 1cm cube tissue volume. The source (inlet) and sink (outlet) are placed along opposite diagonal corners. A steady inflow boundary condition is applied at the inlet ($Q=1.5$mL/min) and a resistance-free boundary condition is applied at the outlet. \textbf{B.} flow split distributions for the entire lattice network \textbf{C.} average pressure drop along vessels for the entire lattice network \textbf{D.} truncated lattice network with near-stagnant flow ($Q<<0.01$ mL/min) vessels removed. Flow preferentially shunts through diagonal vessels between source and sink along the path of least hydraulic resistance \textbf{E.} lattice architecture with 125 vessels for a 1cm cube tissue volume. The source and sink are placed diagonally along the face. The same inlet flow ($Q=1.5$mL/min) and zero-resistance outlet boundary conditions are applied to the source and sink, respectively. \textbf{F.} flow split distributions for the entire network \textbf{G.} average pressure drop along vessels for the entire lattice network \textbf{H.} truncated lattice network with near-stagnant flow ($Q<<0.01$ mL/min) vessels removed. Flow preferentially shunts through diagonal vessels between source and sink along the single face of the lattice network. Vessels towards the opposite face of the source/sink primarily experience near-stagnant flow. 
\clearpage
\begin{figure}[H]
    \centering
    \includegraphics[width=1.1\textwidth]{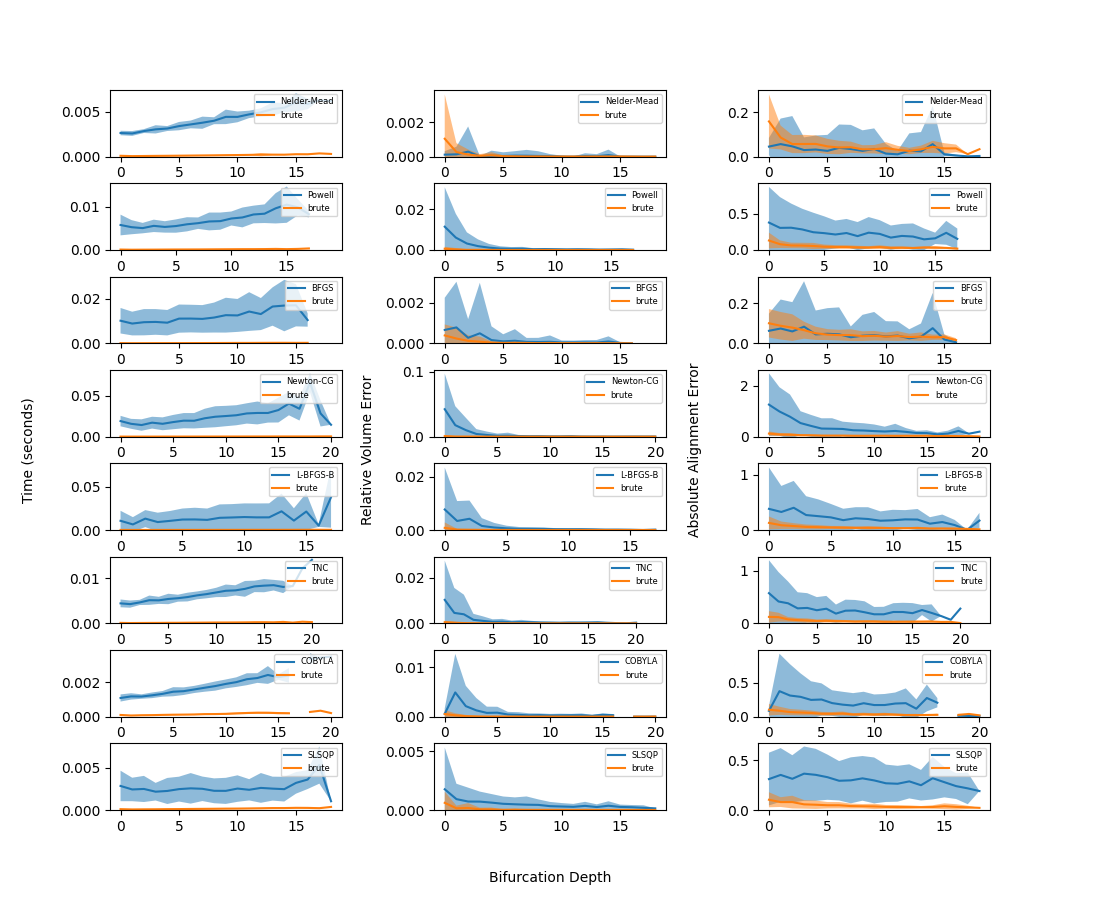}
    \label{S Fig. 3}
\end{figure}
\textbf{Fig. S3. Optimization methods for local bifurcation placement during vascular generation.} Time performance of bifurcation optimization position on a fixed iteration budget (maxiter=20) compared against brute force grid search with a (grid dimension=20) shown in left column. Relative cost function error (blood volume of synthetic tree) compared to ground truth (brute grid search with grid dimension=200) shown in middle column. Absolute distance from the ground truth bifurcation point shown in right column. Shaded area represents $\pm 1$ SD
\clearpage
\begin{figure}[H]
    \centering
    \includegraphics[width=0.8\textwidth]{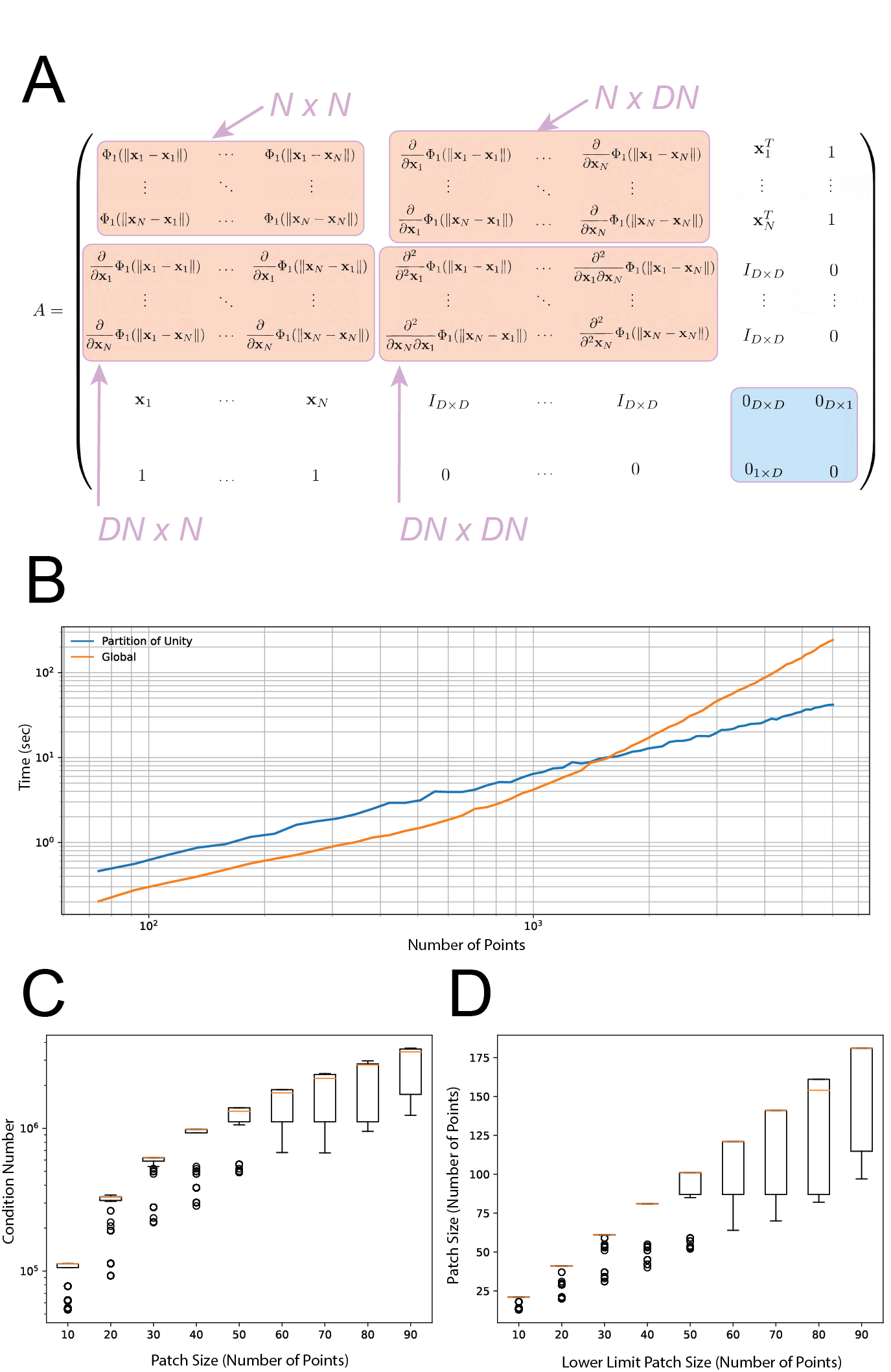}
    \label{S Fig. 4}
\end{figure}
\textbf{Fig. S4. Partial implicit method scaling and numerical stability. A.} Full interpolation matrix for partial implicit patch interpolation. The symmetric matrix requires first and second derivatives of the chosen polyharmonic spline identified by the highlighted submatrices. By default this is the triharmonic ($\phi_{l} = \phi_{1}$). The $(N+1)(D+1)$ rank of this matrix scales depending on the number of interpolating points ($N$) and dimensionality of the domain ($D$). \textbf{B.} because interpolation involves inversion of the full A matrix, global reconstruction schemes (orange) scale poorly above a couple thousand points. Instead the partial implicit method leverages a partition of unity (blue) scheme which creates local interpolation patches of semi-compact support. Such patches limit the upper rank of interpolation matrices and confer more favorable computational scaling. \textbf{C.} given a range of patch sizes we observe a positive relationship with the interpolation matrix condition number with larger patches generally having greater condition numbers \textbf{D.} patch sizes vary with the curvature and distribution of points along the interpolating surface; setting minimum limits to patch sizes ensures that created patches satisfy a normal distribution. Generally lower limits greater than 50 points can ensure minimal outliers. 

\begin{figure}[H]
    \centering
    \includegraphics[width=1\textwidth]{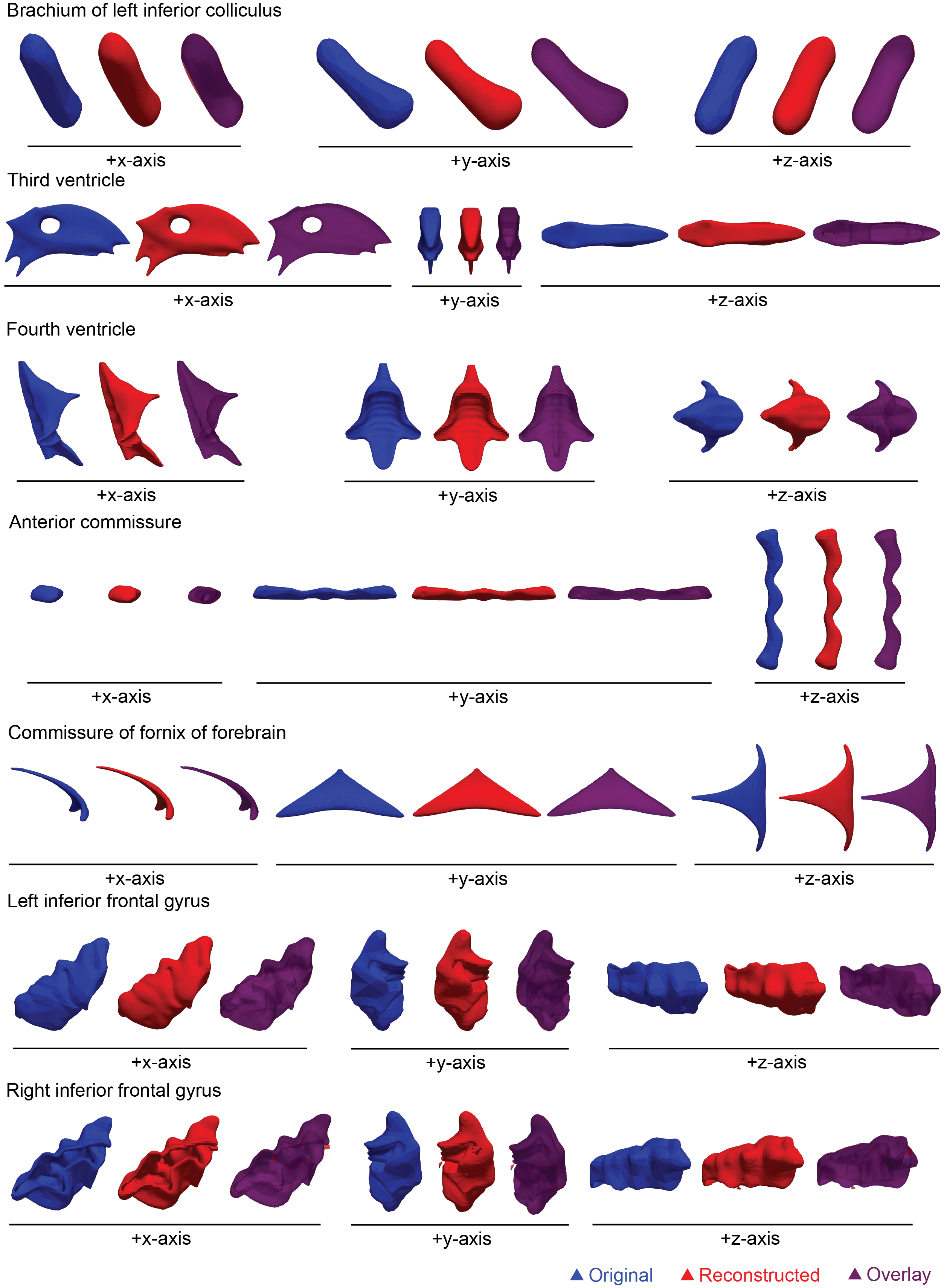}
    \label{S Fig. 5P1}
\end{figure}
\begin{figure}[H]
    \centering
    \includegraphics[width=1\textwidth]{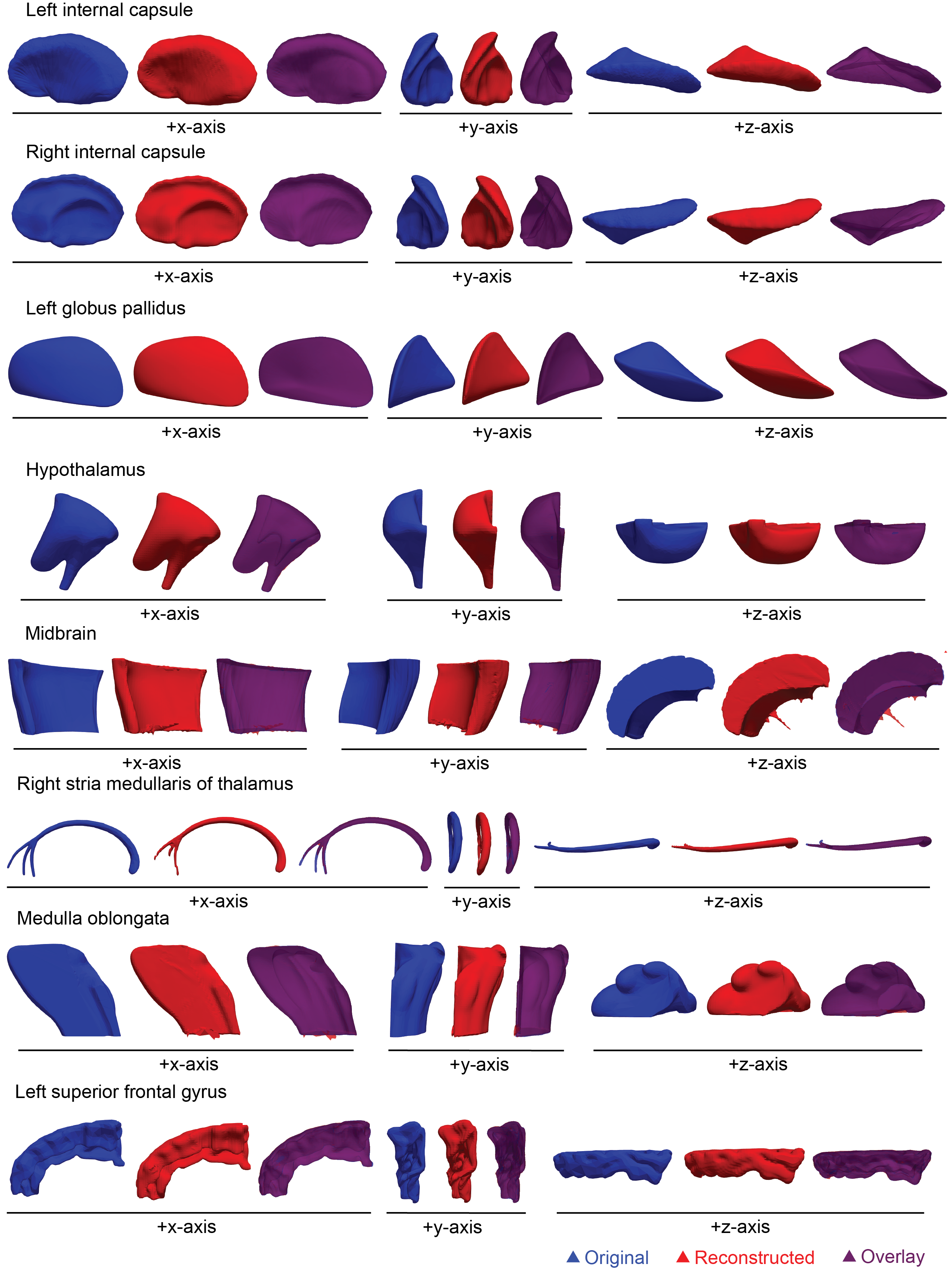}
    \label{S Fig. 5P2}
\end{figure}
\begin{figure}[H]
    \centering
    \includegraphics[width=1\textwidth]{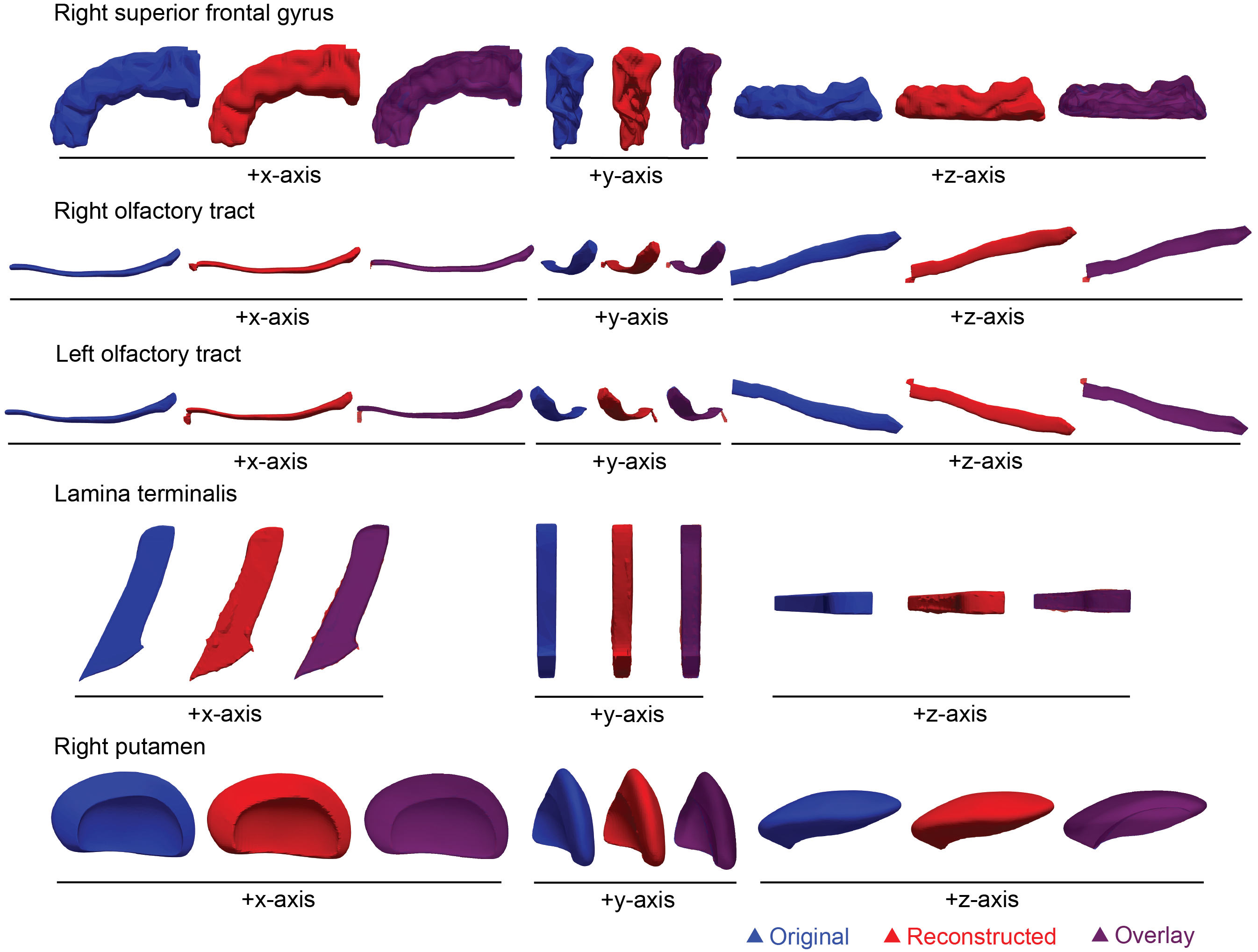}
    \label{S Fig. 5P3}
\end{figure}
\textbf{Fig. S5. Implicit partial volume reconstruction performance.} From the more than 200 anatomic regions of brain tissue obtained from open-source MRI scans and reconstructed with partial implicit methods, twenty reconstructions are shown with views along x, y, and z axes. Original surface meshes (blue) are compared against level-sets along the boundary at $f(\textbf{x}) = 0$ (red) computed from marching cubes. The overlay is also provided with qualitatively similar regions displaying a purple color. Volumetric errors associated with the overlay images are quantified in (Supplement Table \ref{tissue domain reconstruction}).
\clearpage
\begin{figure}[H]
    \centering
    \includegraphics[width=0.8\textwidth]{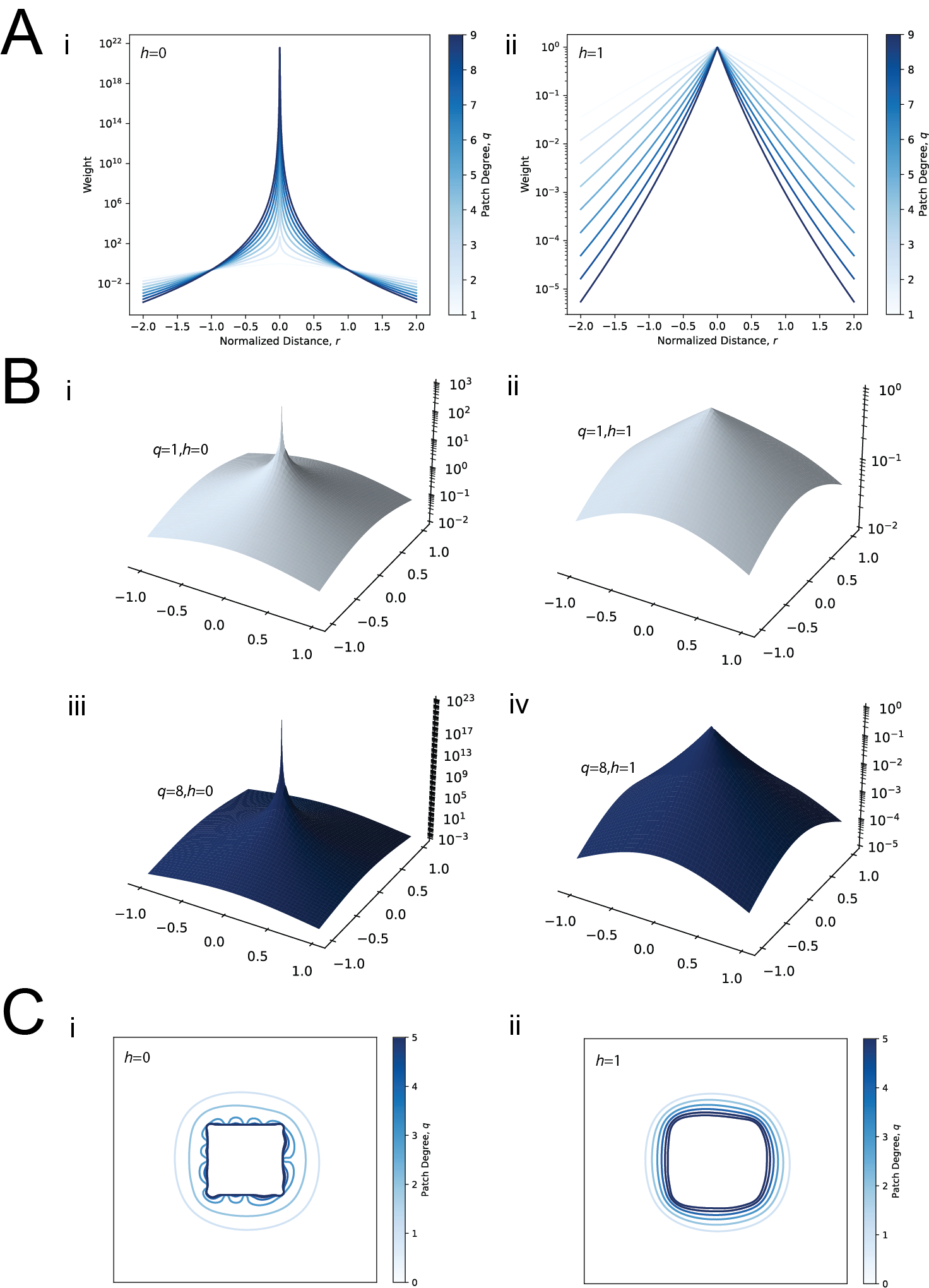}
    \label{S Fig. 6}
\end{figure}
\textbf{Fig. S6. Higher-order Shepard-like weighting functions approximate sharp edges to tissue domains. A.} For the semi-compact weighting function $w_{i} = (1/(h+||\textbf{x}-\textbf{x}_i||))^{q}$ that is used by default in partial implicit methods, values of $h$ and $q$ can be tailored to approximate a Dirac delta function with desired compact support for $h=0$ (i) discontinuities exist at control points of interpolating patches. Values $h > 0$ (ii) present no discontinuities.\textbf{B.} These same functions are shown in 3D to provide intuition for the support of individual patches. With increasing weight degrees $q$, patches provide more compact support.\textbf{C.} greater compact weighting functions are necessary to closely approximate sharp ($\text{C}^{1}$ discontinuous) geometries. This is shown for the edges of a cube where the discontinuous, high-degree weighting function (i) better captures sharp corners than lower-degree weighting functions or non-discontinuous weighting functions (ii).
\clearpage
\clearpage
\begin{figure}[H]
    \centering
    \includegraphics[width=0.8\textwidth]{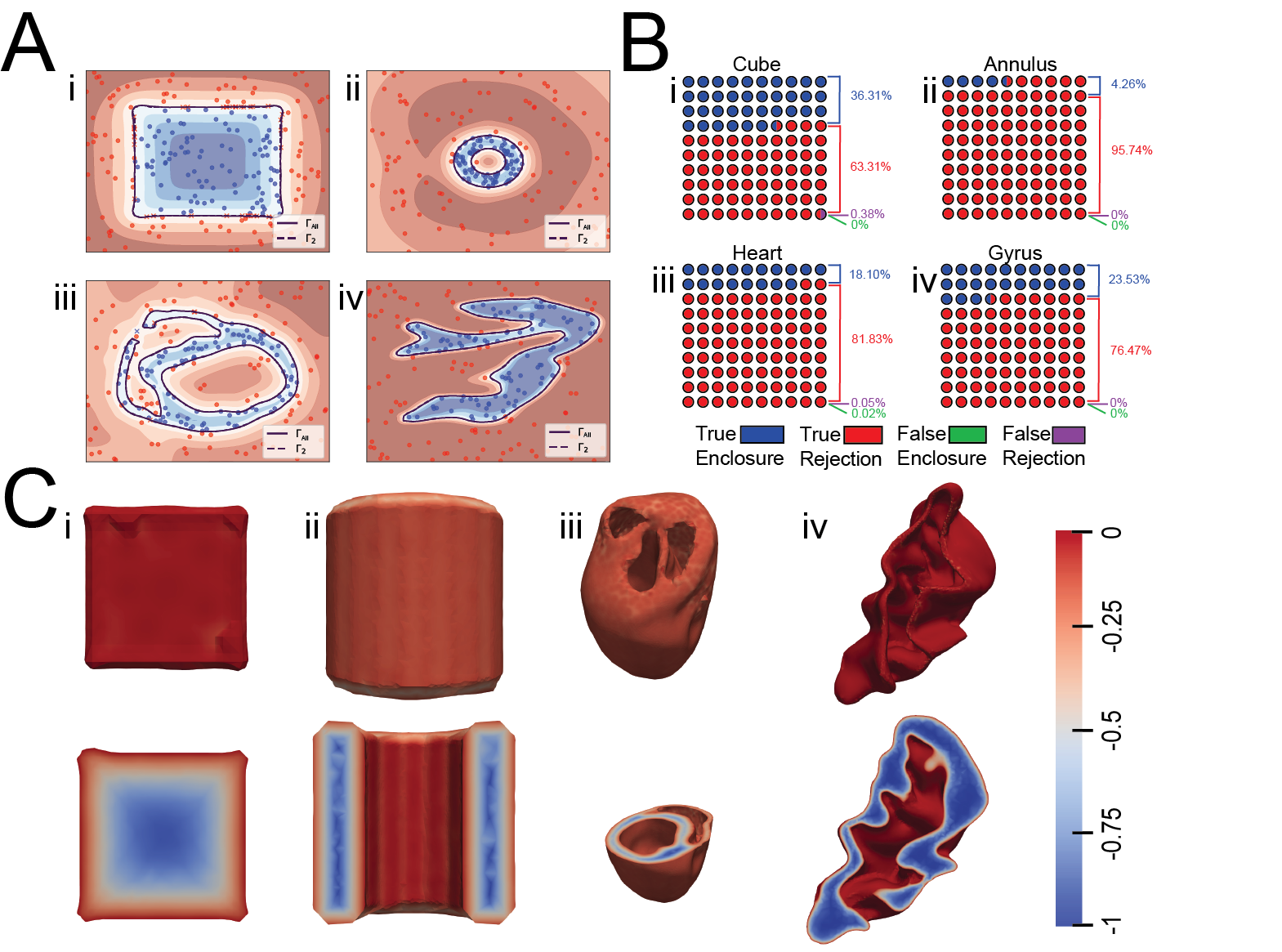}
    \label{S Fig. 7}
\end{figure}
\textbf{Fig. S7. Point enclosure performance for partial implicit volumes A.} 2D slices taken along canonical axes of the partial implicit tissue domains for a cube (i), annulus (ii), biventricle heart (iii), and brain gyrus (iv) are displayed. Red contours depict regions outside the tissue domain while blue contours lie inside the tissue domain. Boundary surfaces $\Gamma$ reconstructed from all $\Gamma_{\text{All}}$ and only two $\Gamma_{2}$ partial implicit patches are shown with solid and dashed lines, respectively, for each geometry. \textbf{B.} the sensitivity and specificity of point enclosure detection via partial implicit methods for a cube (i), annulus (ii), biventricle heart (iii), and brain gyrus (iv). Random points taken along slices shown in panel A are evaluated to be inside or outside the tissue domain using partial implicit methods with mesh-based point-in-polygon evaluations representing the ground truth. Points which are correctly identified as inside the domain are called "true enclosures" while those that are not are "false enclosures." Points which are correctly identified as outside the domain are called "true rejections" while incorrect points are "false rejections." True and false enclosures are displayed as blue dots and crosses, respectively; true and false rejections are displayed as red dots and crosses, respectively within panel A. \textbf{C.} 3D meshes reconstructed from the zero level-set surface ($L_{0}$) obtained via a marching cube algorithm display values of partial implicit functions throughout volume. Interior values are seen after slicing the 3D meshes for a cube (i), annulus (ii), biventricle heart (iii), and brain gyrus (iv). Ranges for interior values span [0,-1] inclusively. 
\clearpage
\begin{figure}[H]
    \centering
    \includegraphics[width=0.7\textwidth]{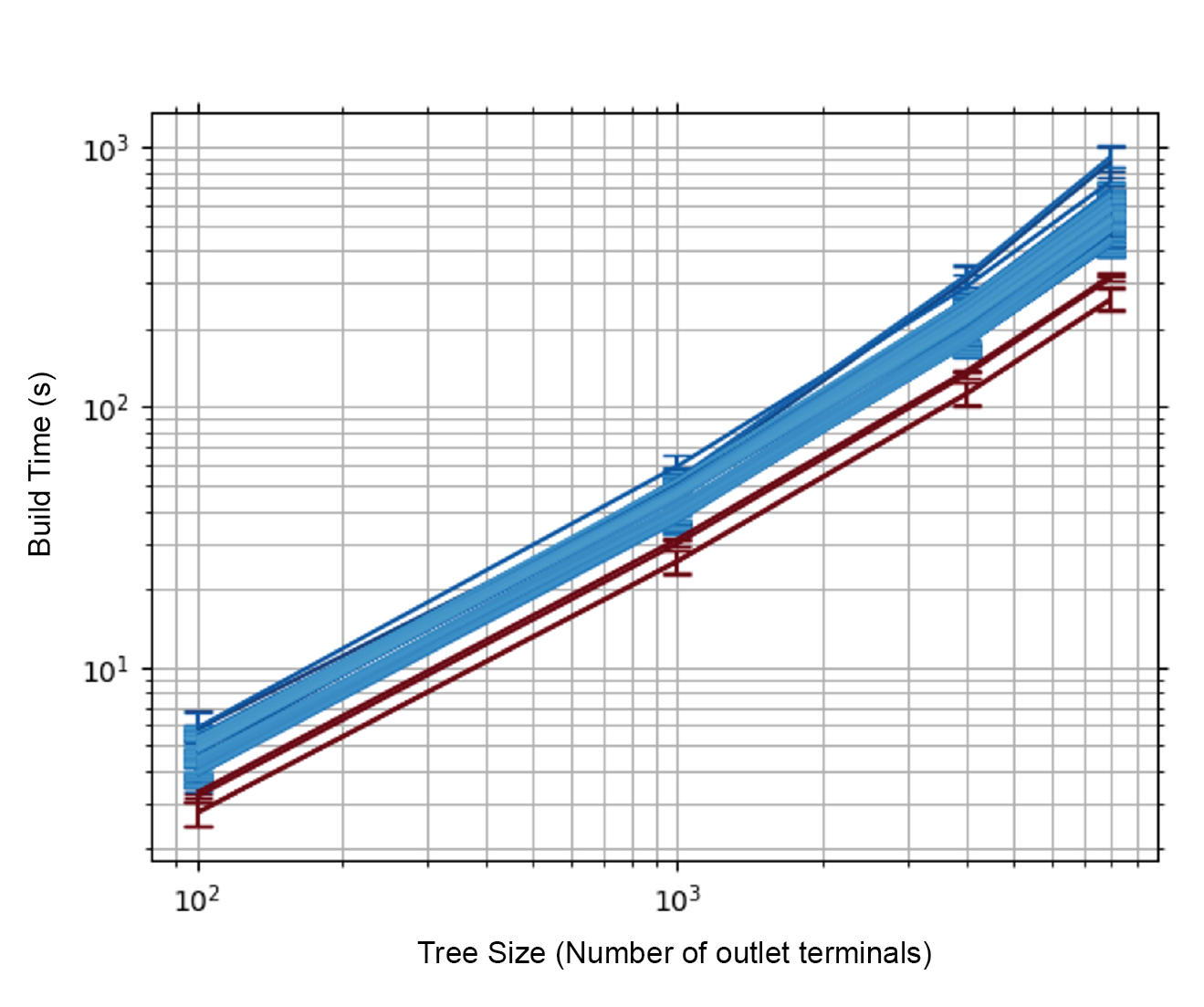}
    \label{S Fig. 8. }
\end{figure}
\textbf{Fig. S8. Estimated build times for synthetic vasculature across a large set of engineering and anatomic tissue domains.} Build times for synthetic vascular tree generation are made for 10 engineering and 208 anatomic models for trees with 100, 1000, 4000, and 8000 terminal segments (error bars shown). Blue curves represent tissue domains which are non-convex; red curves represent convex tissue domains. For each tissue domain, 100 synthetic vascular trees are generated to obtain time performance data. Error bars are $\pm$ 1SD. 
\clearpage
\begin{figure}[H]
    \centering
    \includegraphics[width=0.8\textwidth]{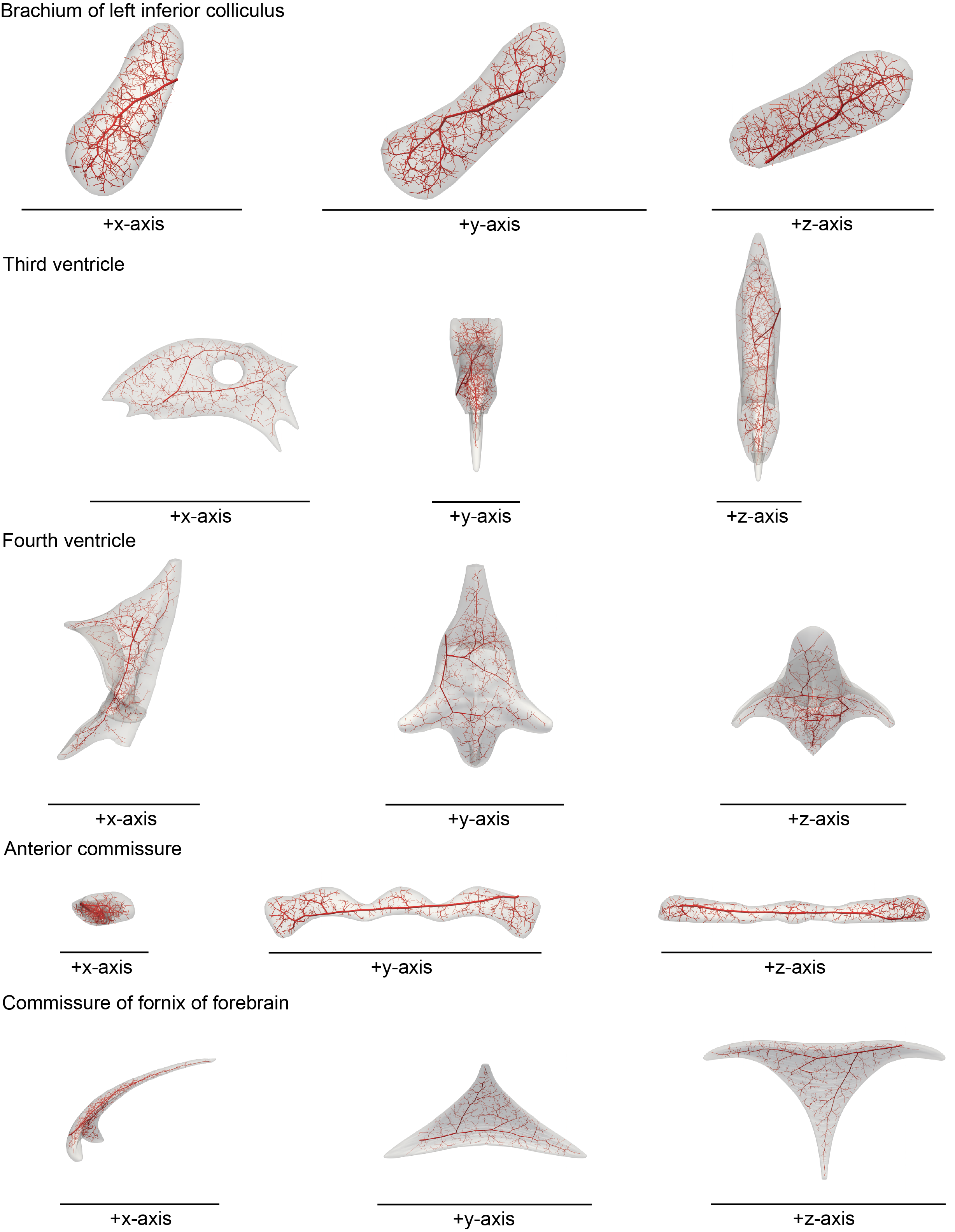}
    \label{S Fig. 9P1}
\end{figure}
\begin{figure}[H]
    \centering
    \includegraphics[width=0.8\textwidth]{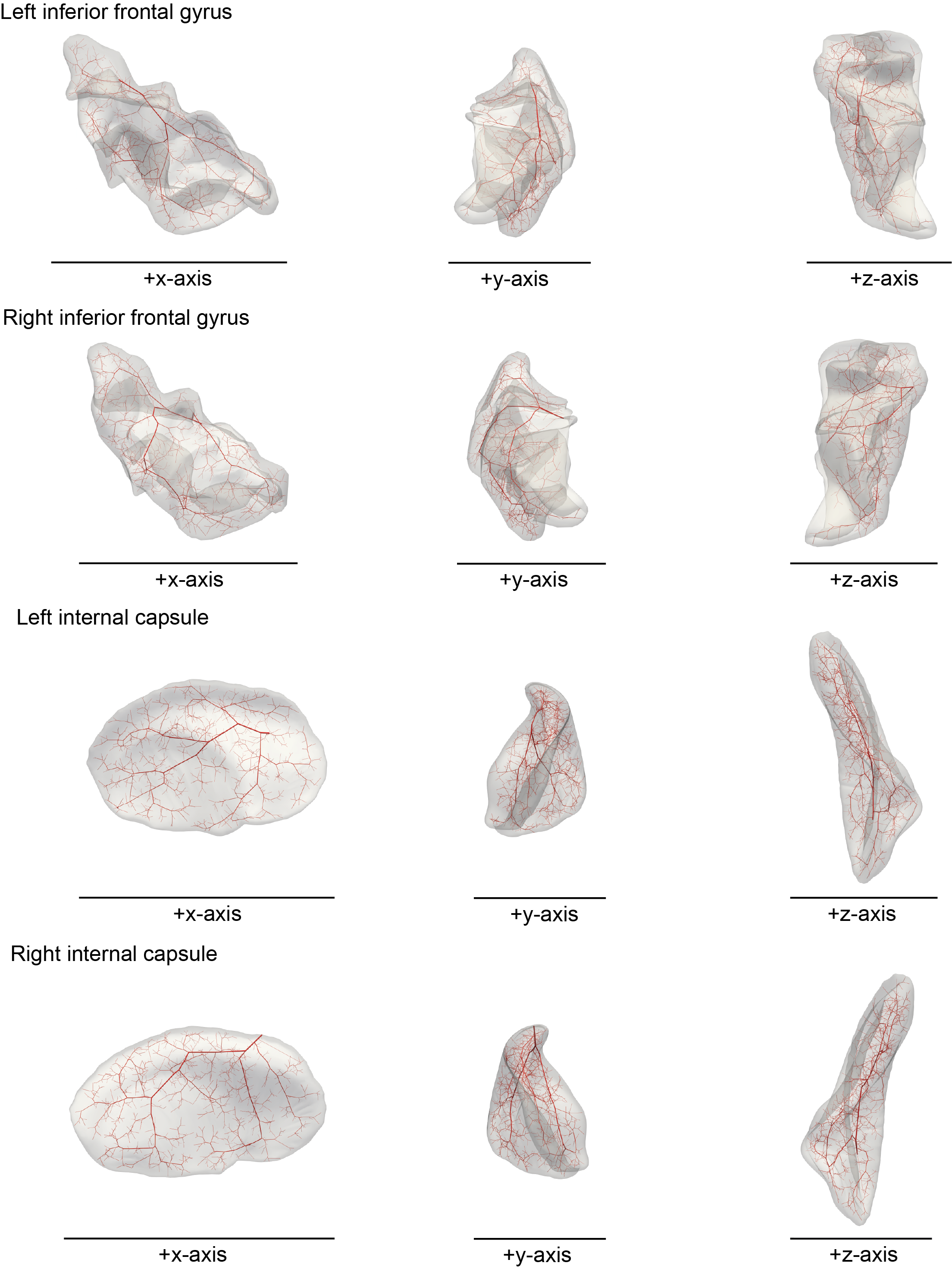}
    \label{S Fig. 9P2}
\end{figure}
\begin{figure}[H]
    \centering
    \includegraphics[width=0.8\textwidth]{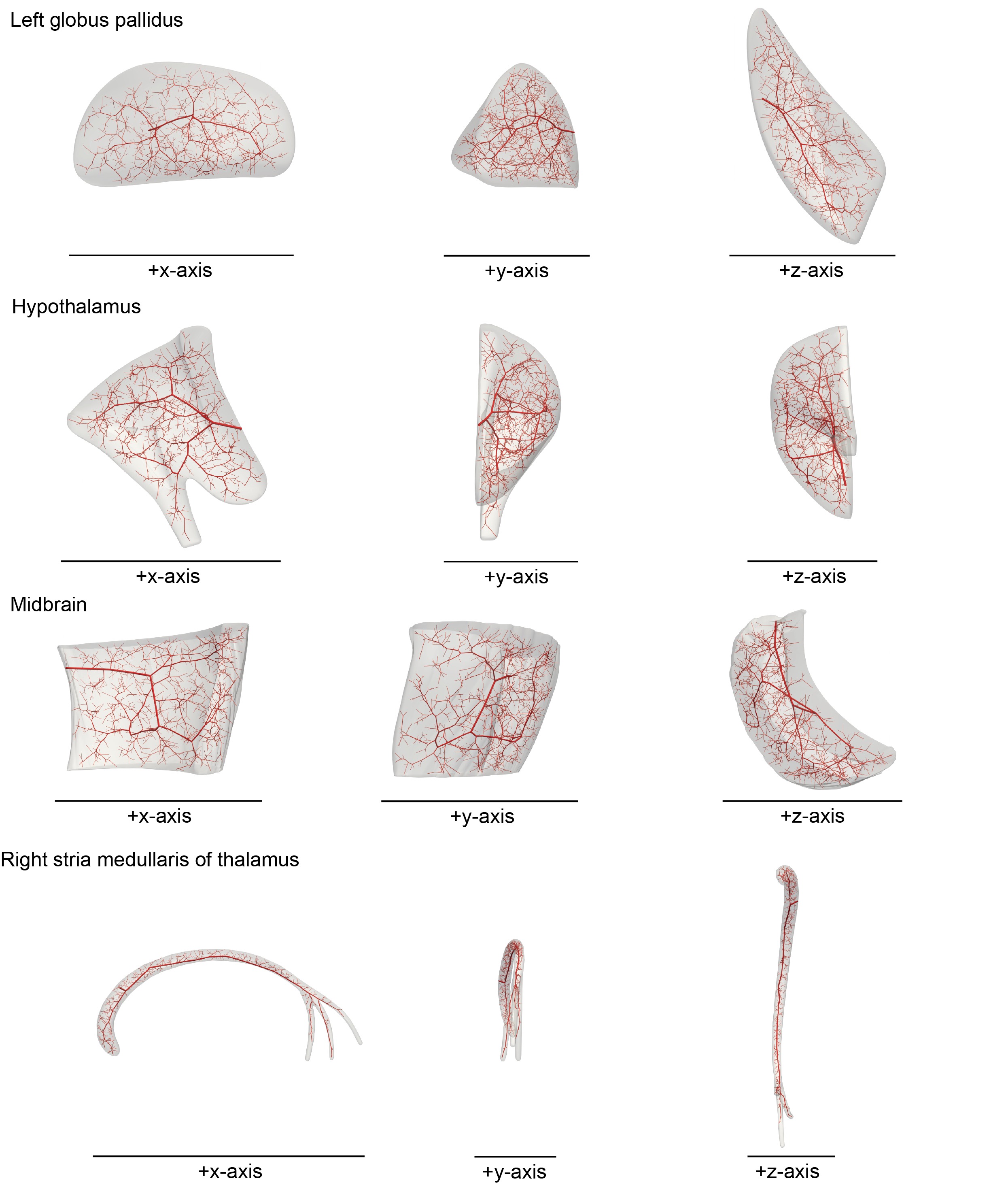}
    \label{S Fig. 9P3}
\end{figure}
\begin{figure}[H]
    \centering
    \includegraphics[width=0.8\textwidth]{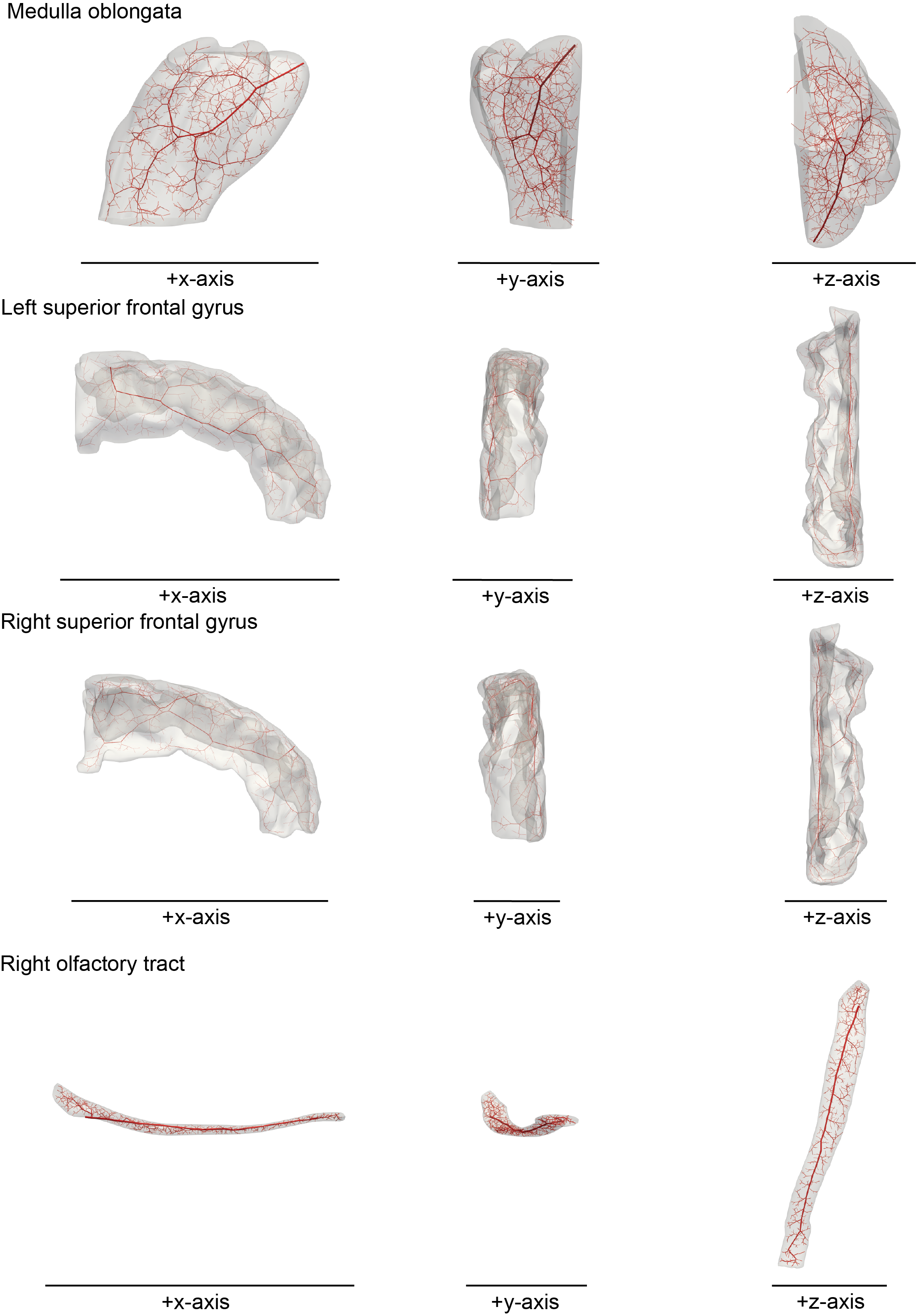}
    \label{S Fig. 9P4}
\end{figure}
\begin{figure}[H]
    \centering
    \includegraphics[width=0.8\textwidth]{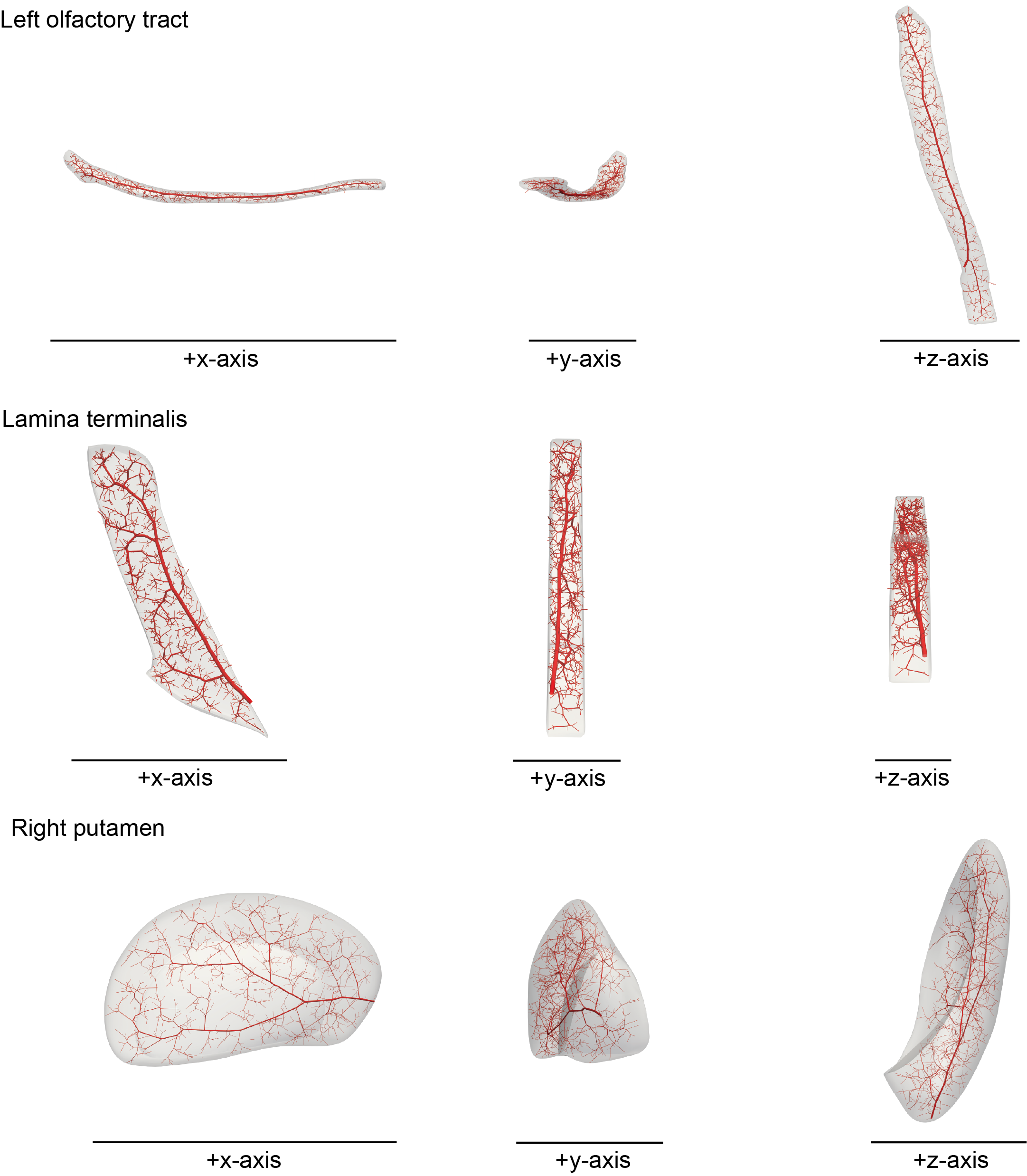}
    \label{S Fig. 9P5}
\end{figure}
\clearpage
\textbf{Fig. S9. Synthetic vascular generation in anatomic regions of brain tissue.} For twenty of the more than 200 anatomic regions of brain tissue, synthetic vascular trees with 1000 terminal segments are generated and visualized along the x, y, and z axes. Associated time performance for each domain is given in (Supplement Table \ref{tissue domain vascularization}). For these tissue domains, an open-source benchmark CCO algorithm is also run under the same build parameters; benchmark times are provided in the same table. 
\clearpage
\newpage
\textbf{Movie S1. Synthetic vascularization of arbitrary tissue domains.} Video shows 3D camera orbits about the centroids of a vascularized cube, annulus, biventricle heart, and brain gyrus. Each synthetic vascular network has 1000 terminal outlets and is fully contained within its respective tissue domain (red transparent shells). Tissue domain convexity is also given where convexity = Volume/Volume$_{\text{Delaunay}}$.
\\
\textbf{Movie S2. Pressure pulse propagation from time-varying flow profiles.} Video shows reduced-order 0D model simulations for a cube (i), annulus (ii), biventricle heart (iii), and brain gyrus (iv) with 1000 terminal synthetic vascular networks. A time-varying inflow boundary condition is applied at the root of each vascular tree. Zero resistance boundary conditions are applied uniformly across all outlets. Displayed are results for average normalized cross-sectional pressure over time. Simulations are ran for 6 cardiac cycles with 20 time points per cycle. Normalized inflow rate is shown for the cardiac cycles to the left of the pressure simulations. Speed 1x. 
\\
\textbf{Movie S3. Perfusion of simple planar synthetic vascular FRESH network.} Video shows steady flow ($Q=0.25$mL/min) perfusion of simple planar vascular network with black dye. Speed 1x. 

\bibliography{supplement}

\bibliographystyle{Science}

\clearpage